\documentclass[sigconf,authorversion,nonacm]{acmart}

\usepackage{svg}

\usepackage{listings,multicol}
\usepackage{xcolor}
\usepackage{algpseudocode}
\usepackage{algorithm}
\algnewcommand{\LineComment}[1]{\State \(\triangleright\) #1}

\usepackage{soul}

\usepackage[title]{appendix}
\usepackage{caption}
\usepackage{subcaption}
\usepackage{tikz}
\usetikzlibrary{shapes.geometric, arrows}
\usetikzlibrary{positioning,fit,calc}
\tikzstyle{startstop} = [rectangle, rounded corners, minimum width=3cm, minimum height=1cm,text centered, draw=black, ]
\tikzstyle{io} = [trapezium, trapezium left angle=70, trapezium right angle=110, minimum width=3cm, minimum height=1cm, text centered, draw=black]
\tikzstyle{process} = [rectangle, minimum width=3cm, minimum height=1cm, text centered, draw=black]
\tikzstyle{decision} = [diamond, minimum width=3cm, minimum height=1cm, text centered, draw=black, fill=green!30]
\tikzstyle{arrow} = [thick,->,>=stealth]

\definecolor{codegreen}{rgb}{0,0.6,0}
\definecolor{codegray}{rgb}{0.5,0.5,0.5}
\definecolor{codepurple}{rgb}{0.58,0,0.82}
\definecolor{backcolour}{rgb}{1,1,1}

\lstdefinestyle{mystyle}{
  backgroundcolor=\color{backcolour},   
  commentstyle=\color{codegreen},
  keywordstyle=\color{magenta},
  numberstyle=\tiny\color{codegray},
  stringstyle=\color{codepurple},
  breaklines=true,
  numbers=left,                    
  numbersep=2pt,                  
  showspaces=false,                
  showstringspaces=false,
  showtabs=false,  
  tabsize=1,
  basicstyle=\small
}

\AtBeginDocument{%
  }

\setcopyright{acmlicensed}
\copyrightyear{2024}
\acmYear{2024}
\acmDOI{XXXXXXX.XXXXXXX}

\begin{document}
\title{Efficient Dataframe Systems: Lazy Fat Pandas on a Diet}

\author{Bhushan Pal Singh}

\affiliation{%
  \institution{IIT Bombay}
}
\email{singhbhushan@cse.iitb.ac.in}

\author{Priyesh Kumar}
\affiliation{
 \institution{IIT Bombay$^\#$ \thanks{$^\#$ Current Affiliation: Dream11} }
}
\email{priyeshkumar9875@gmail.com}

\author{Chiranmoy Bhattacharya}
\affiliation{%
 \institution{IIT Bombay$^*$ {\thanks{$^*$ Current Affiliation: Fujitsu, India}}}
}
\email{chiranmoy358@gmail.com}

\author{S. Sudarshan}
\affiliation{%
  \institution{IIT Bombay}
  }
\email{sudarsha@cse.iitb.ac.in}



\begin{abstract}
Pandas is widely used for data science applications, but users often run into problems when datasets are larger than memory.  There are several frameworks based on lazy evaluation that handle large datasets, but the programs have to be rewritten to suit the framework, and the presence of multiple frameworks complicates the life of a programmer.

In this paper we present a framework that allows programmers to code in plain Pandas; with just two
lines of code changed by the user, our system optimizes the program using a combination of just-in-time static analysis, and runtime optimization based on a lazy dataframe wrapper framework.  
Moreover, our system allows the programmer to choose the backend. It works seamlessly with Pandas, Dask, and Modin, allowing the choice of the best-suited backend for an application based on factors such as data size.  Performance results on a variety of programs show the benefits of our framework.
\end{abstract}


\maketitle

\newcommand{\eat}[1]{}
\newcommand{\fullversion}[1]{}
\newcommand{\shortversion}[1]{#1}
\newcommand{\SCIRPy}{SCIRPy}

\newcommand{\inred}[1]{\textcolor{red}{#1}}

\makeatletter 
\let\c@table\c@figure
\let\c@lstlisting\c@figure
\let\c@algorithm\c@figure
\makeatother

\section{Introduction}
\label{sec:intro}

Python is widely used for data science applications, and in particular dataframe-based libraries and frameworks have become the default model for many data science applications. Pandas is the most popular framework among these and is the tool of choice for applications that use smaller datasets that fit in the available system memory.  To address the needs of applications that have larger datasets that do not fit in memory, several scalable frameworks have been created, such as  Dask ~\cite{rocklin2015dask}, Modin ~\cite{modin}, PySpark~\cite{pysparkpandas}, Magpie~\cite{Magpie}, among others. 
Some frameworks such as Dask and Vaex are lazy evaluation frameworks that create a task graph lazily, optimize it, and then execute it when results are needed. Others like Modin support eager evaluation. 
\fullversion{
All these make use of multiple CPU cores as against Pandas which is inherently single-threaded. 
}


Many users develop their applications using Pandas, and test them on small datasets; performance issues including out-of-memory issues, are not obvious until much larger datasets are used.  Even if production datasets fit in memory at a point in time, the growth of data sizes often causes problems at a later point in time.

While users can avoid these problems by using scalable frameworks, there are several challenges.  Frameworks based on lazy evaluation can speed up evaluation by using optimizations of the task graph, but require users to write the code in a manner suitable for lazy evaluation.  For example, 
explicit forcing of computation is required
before printing a lazily computed dataframe. 
Availability of a large number of frameworks coupled with variations in their APIs, including lack of support for some Pandas features, further complicate the work of the user.  Finally, which framework is optimal for an application depends on factors such as data size.   Thus, a significant burden is placed on the programmer to choose and use an appropriate framework.  This may be contrasted with the declarative specifications and automated optimization supported by languages such as SQL.

In this paper, we propose an efficient and scalable dataframe system, which we call \textit{Lazy Fat Pandas} (\textit{LaFP}), which addresses the above-mentioned challenges faced by the data science user community.

\begin{figure}
\centering
\includegraphics[width=3.4in,height=2.625in,bb=0 0 539 366]{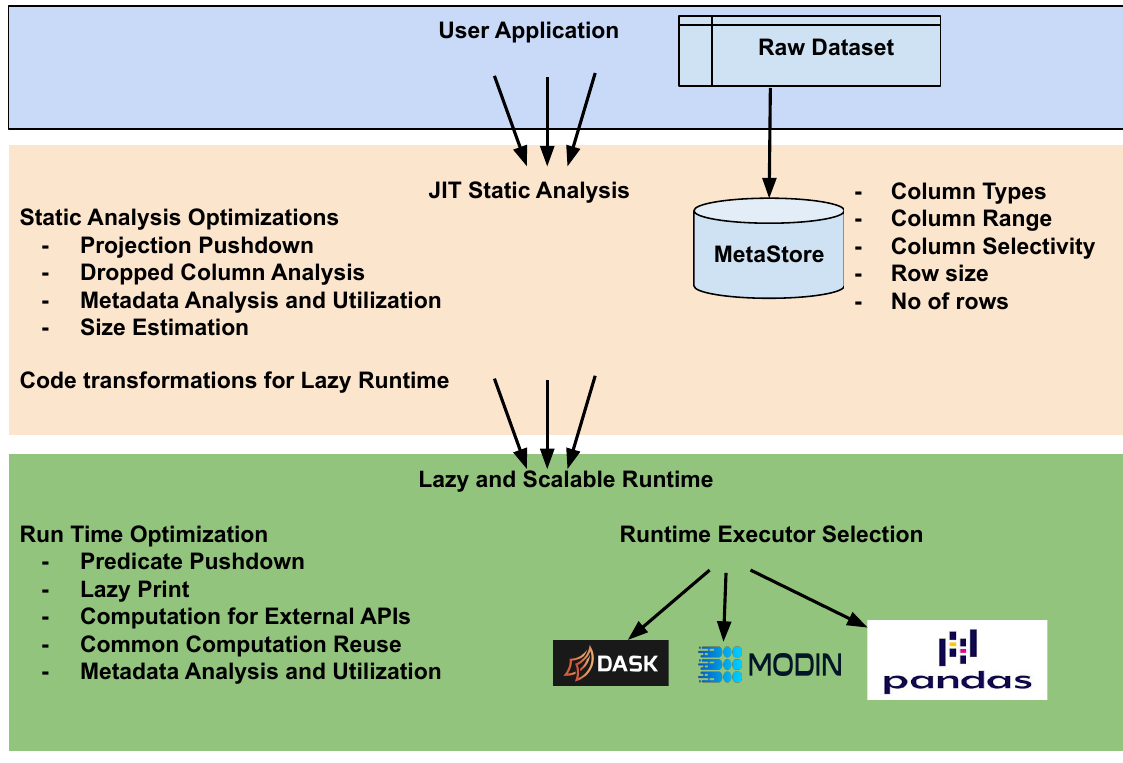}
\caption{Overview of Our Lazy Fat Pandas System}
\label{fig:lfp_overview}
\end{figure}

Figure~\ref{fig:lfp_overview} provides an overview of our system. 
Users write programs in Pandas, and can then use our optimization framework with a mere change in an import statement and one other function call.
Using LaFP, the users need not worry about the details of different back-end executors, or their APIs, or even about lazy evaluation; users can write their application in the plain Pandas APIs, and leave the task of optimization to LaFP. 

Our system combines optimization using Just-in-Time (JIT) static analysis and rewriting of Pandas programs, coupled with a lazy runtime that performs a variety of run-time optimizations, allowing optimization to combine the best of static and run-time optimizations.   


The specific technical contributions of this paper include the following:
\begin{enumerate}

\item  Our lazy runtime wrapper, \textit{Lazy Fat Pandas} (\textit{LaFP}), allows substitution of invocations of Pandas API calls by invocations of lazy versions of the same calls in LaFP.  Lazy evaluation allows the construction of a DAG of operators which can be optimized before execution, when results are needed.  

LaFP supports the choice of different back-end executors (currently, Dask and Modin in addition to Pandas), and executes the operator DAG on the chosen back-end framework.  
When a lazy back-end is used, LaFP creates an operator DAG in the backend framework, whereas if an eager back-end is chosen LaFP invokes individual operations in sequence to get the required result.

LaFP performs a number of runtime optimizations, motivated by transformations used in query optimizations, on these programs. 
Examples include Predicate Push Down, Forced Computation for External Function Modules, and Common Computation Reuse.   Lazy frameworks like Dask can subsequently perform their own optimizations.

LaFP also provides novel lazy implementations of operations that are not part of the Pandas framework, in particular Lazy Print.  Lazy Print helps postpone forcing of evaluation beyond the occurrences of print statements, and can improve performance.

\item  We present an optimization architecture based on static analysis, to 
perform source-to-source transformation of Python programs.
The Python source programs is first converted to the \SCIRPy\  intermediate representation, which we have created, which is compatible with the Soot static analysis framework \cite{Soot}.  
Rewriting is performed based on static analysis, and the IR is converted back to Python for execution.  
Our optimizer also rewrites Pandas function invocations to optimally use our Lazy Fat Pandas framework, which enables a number of run-time optimizations.  

\item We define a novel static analysis and optimization methodology that utilizes \textit{{\ul{just-in-time (JIT) static analysis}}}.  Our approach requires only the addition of a single function call to the program, which uses reflection to find and rewrite the source code of the program, and replace the execution of the original program by the execution of the optimized rewritten program.   No changes are required to the outer-level systems that invoke the Python programs, greatly simplifying the task of deploying the optimizations.

\item  We present an optimization architecture that \textit{\ul{combines optimization based on static analysis with run-time optimization}}, to get the best of two worlds.  Static analysis allows our optimizer to look ahead to predict which data frames are live, and what parts of the dataframe (based on column selections) will get used later in the program, beyond what a run-time optimizer based on lazy frameworks can do.  The run-time optimizer based on the lazy framework allows us to implement a variety of optimizations akin to a database query optimizer using information that is not available at compile time. 

We also implement a system for gathering metadata (types and statistics) about data sets for use in optimization.



\item Our LaFP library supports {diverse back-ends}, such as Pandas, Dask and Modin.

The backends differ in the support for specific Pandas API functionality, but more fundamentally  Dask does not support ordering of rows in a dataframe.
If a chosen back-end does not support a specific Pandas API functionality, LaFP is able to convert data from the back-end representation back to Pandas, to execute the original Pandas function.  Thus the user need not worry about the differences between back-ends (beyond the lack of row ordering in Dask), or their specific limitations.  
The choice of backend is currently made as a program configuration, but we 
are working on automating the choice in future, based on factors such as size of the datasets and row order dependence.

\item We have implemented the LaFP framework, along with static and run-time optimizations.  Our performance studies show that our methodology can significantly improve the performance of such programs with {up to 19X speedup of execution time}, and allow execution of 9 out of 10 programs on a 12.6 GB dataset, where only 2 could run using Pandas.
\end{enumerate}

While there are several systems that support lazy runtimes with optimization, such as Dask and Pandas on Spark, and a few systems that perform code rewriting (see Section~\ref{sec:relwork} for details), to our knowledge, no other system performs such combined optimization.  Our performance study illustrates the benefits of such combined optimization.

The rest of the paper is organized as follows.
Section~\ref{sec:arch} describes our system architecture.  Section~\ref{opt} describes the optimizations implemented in our system.
Section~\ref{sec:relwork} describes related work.
Section~\ref{sec:perf} presents a performance study showing the benefits of our approach, while
Section~\ref{sec:concl} concludes the paper.

\fullversion{
(An initial version of the intermediate representation, along with initial versions of column selection and metadata analysis optimizations based on static analysis were described in a short 4 page paper \cite{anon} (citation anonymised), but all the other contributions mentioned above are novel to this paper.)
}

\section {Architecture}
\label{sec:arch}

Figure ~\ref{fig:lfp_overview} shows an overview of our proposed framework and the static and runtime optimizations implemented in our system.

\begin{figure}
\centering
\includegraphics[bb=0 0 629 110,width=3.5in,height=0.75in]{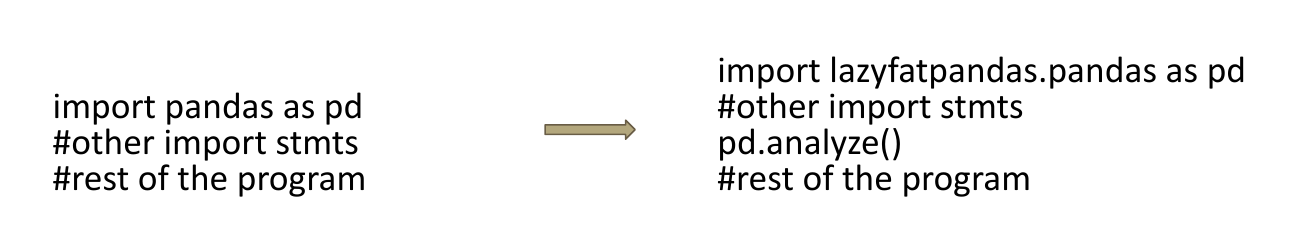}
\caption{Code changes for Using Lazy Fat Pandas Framework}
\label{fig:jit}
\end{figure}

Users can run existing Pandas-based programs using our Lazy Fat Pandas (LaFP) framework with minimal changes.  As highlighted in Figure~\ref{fig:jit}, the user just replaces an import of pandas by an import of our lazyfatpandas library, and adds a call to pd.analyze().  
Our lazyfatpandas API supports the Pandas API.\footnote{Although we have not yet implemented all the API functions, the bulk of the widely used API functionality is supported, and we are adding support for further API calls.} 

\fullversion{
The user program is initially optimized using Just-in-Time (JIT) static analysis and optimization phase as discussed in Section ~\ref{jit}. The output of this phase is a statically optimized program. This phase also builds metadata information for datasets available at compile time. An initial assessment of the cost of the program execution can also be made based on the input datasets used in the program. 

The optimized program is then executed in a lazy fashion as discussed in Section \ref{lazyscalableruntime}. Rather than performing the operation eagerly, a task graph is constructed for various LaFP API calls. This task graph is then optimized for run-time optimizations. The optimized task graph is then executed to run program optimally. LaFP can use a number of executors for execution of the optimized task graph. LaFP currently supports Pandas, Dask, and Modin frameworks as executors. It can be extended to support other Scalable frameworks. 
}

\subsection{Static Analysis and Rewriting Framework}
\label{ssec:opt_framework}

\fullversion{
\begin{figure}[bt]
\centering
\fullversion{\includegraphics[width=3in,height=2.0in]{figures/flow.jpg}}
\shortversion{\includegraphics[width=3in,height=2.0in]{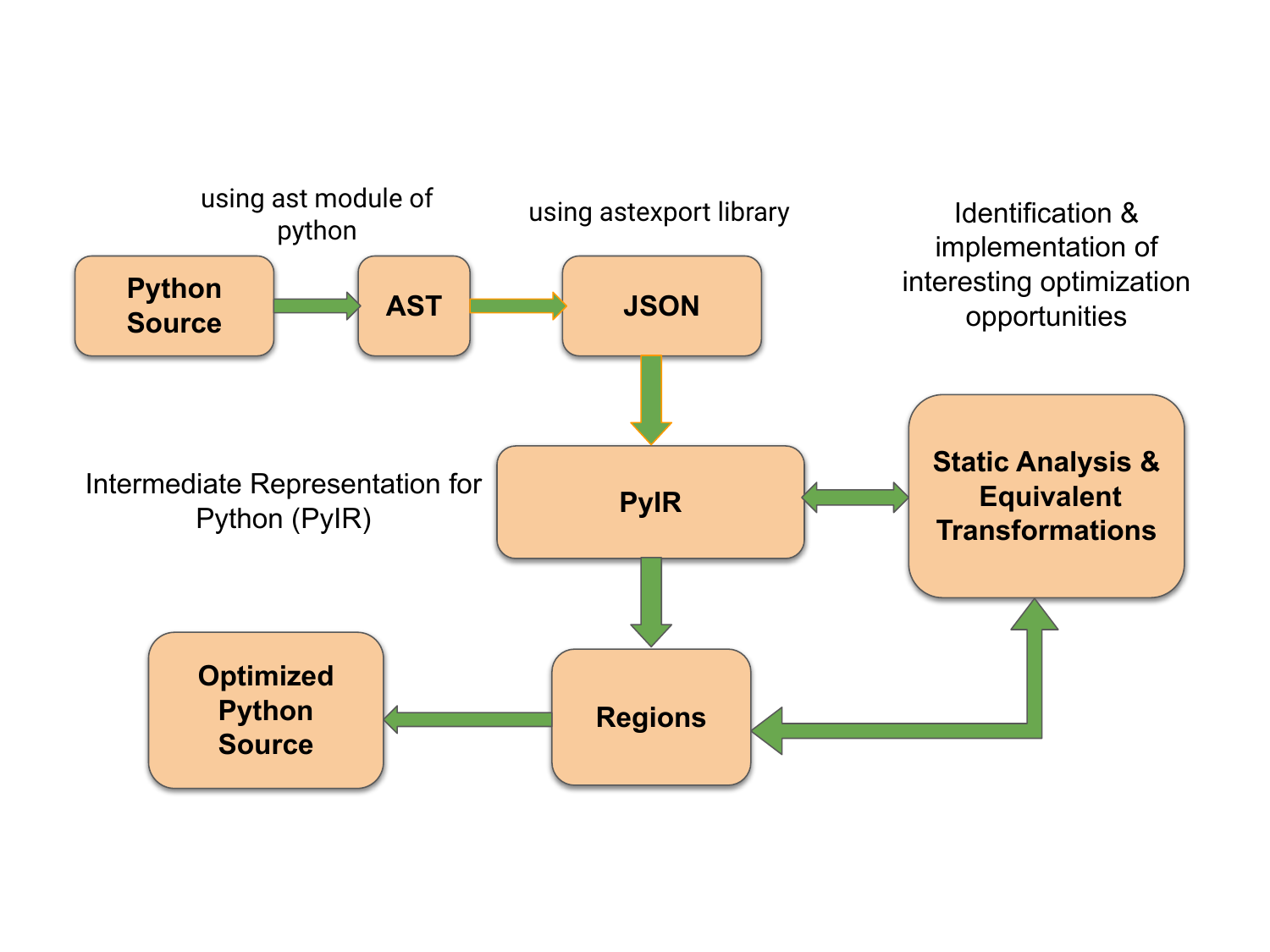}}
\caption{Steps in Static Analysis Based Optimization}
\label{fig:SCIRPy}
\end{figure}
}

Our framework first converts programs to a lower-level internal representation (IR), then performs static analysis, and based on the static analysis results, it
rewrites the program to optimize evaluation.

We use Soot \cite{Soot}, a Java-based framework for static analysis of Python programs.  Soot natively supports analysis of Java byte code by translating it to one of several supported IRs. 
We have extended Jimple IR of Soot to represent Python based programs. We call this extended IR \SCIRPy.  \SCIRPy\ is explained in detail in Section ~\ref{ir_label}.

\fullversion{
\inred{Ensure this figure is not a direct copy of DBPL figure}
The steps carried out by our system during optimization of a program are shown in Figure ~\ref{fig:SCIRPy}. 
}

Python source is parsed into an abstract syntax tree (AST) using an existing parser (written in Python), and the AST is then translated into a JSON (JavaScript Object Notation) object. This JSON object is then read by a Java program that creates a \SCIRPy\ representation of the program.   

Static analysis is then carried out using Soot on the \SCIRPy, and based on the analysis, optimizations are carried out by transforming the IR, as discussed in Section ~\ref{comopt}.
Finally, the optimized IR is converted back to Python code, which is subsequently executed.

We note that static analysis cannot handle dynamically generated code that is executed using the exec() function, and
static analysis of Python, has other challenges such as not knowing which overloaded function is being
called due to lack of static typing. However, Pandas applications typically do not use features that cause such issues,
and our analysis is conservative.

\fullversion{
\begin{algorithm}[bt]
\caption{Python to \SCIRPy}
\label{algo:source_to_scirpy}
\begin{algorithmic}
\Procedure{\textbf{python\_to\_\SCIRPy(source\_file)}}{}

\LineComment{Convert source\_file to json }
\State \textbf{json $\leftarrow$ Transform source\_file to ast and dump JSON }
\LineComment {Read statements from JSON Objects}
\State \textbf{stmts $\leftarrow$ parseJSON(json)}
\LineComment{Transform list of statement to \SCIRPy }
\State \textbf{sootClasses $\leftarrow$ getSootClasses(stmts)}
\State \textbf{mainClass$\leftarrow$ getMainClasses(sootClasses)}
\State \textbf{mainMethod$\leftarrow$ getMainMethod(mainClass)}
\State \textbf{build\_\SCIRPy(mainMethod,mainClass,sootClasses)}
\EndProcedure

\end{algorithmic}
\end{algorithm}
}

\subsection{Intermediate Representation}
\label{ir_label}


Our Soot-based intermediate representation, 
\SCIRPy, is  compatible with the Python Abstract Syntax Tree (AST) representation. This allows the representation of Python's code statements in Soot-compatible units (statements). 
Most of the representations in the IR, such as if statements, assignment statements, etc.\ extend constructs from the Jimple IR of Soot, making it easy to use the built-in analysis functionality of Soot. 
\SCIRPy\ maintains compatibility with Soot by extending Soot's constructs like Soot's class, method, and body for Python.\footnote{
Extending the IR to handle exceptions is a part of future work. 
}
\fullversion{
The body contains three chains for units (i.e. statements), traps (exceptions) and locals (i.e. variables).
}


The control flow graph (CFG) is a representation of the flow of control within a program~\cite{muchnik1997}. It represents various paths that the program execution may take. It is a directed graph in which every node represents a Basic Block (BB). A basic block is a sequential fragment of code without any branch or loop. 
CFGs are generally constructed on intermediate representation (IR). Our implementation constructs CFG from \SCIRPy. 

Data flow analysis~\cite{dfabook} is a technique to identify how a program or a method manipulates its data, and is performed on the control flow graph. 
We define and use live attribute analysis (Section ~\ref{opt:DS}) and live dataframe analysis (Section~\ref{al:ccr}), which are based on data flow analysis.


\fullversion{
As shown in Figure ~\ref{algo:source_to_scirpy}, Python program is transformed into its AST which is transformed into JSON notation where each statement is represented by a JSON object. This JSON notation is parsed into SCIRPy IR. SCIRPy IR extends multiple constructs from Soot to represent Python programs.
}

Based on the static analysis results, the program is rewritten to optimize execution, and then converted back to Python for execution.
To convert programs from \SCIRPy\ back to Python, we
first convert the CFG-based IR to another intermediate representation known as program regions, which represent the hierarchical structure of block-structured programs.  
Regions could be basic block regions, loop regions, branch (if-then-else) regions, or sequential regions, each of which could contain subregions.
Creating regions from the graph based \SCIRPy\ representation is done using techniques described in \cite{hechtullman1972}, which are also used in \cite{equivsql2016}.  The region-based representation
is then translated to Python.  
\fullversion{
In future we plan to 
also explore optimizations that can be done in the region representation on the lines of \cite{equivsql2016}.
}

\fullversion{
A program region has a single entry and single exit point.  A region can be:
\begin{itemize}
    \item a branch region (BR) which is composed of an if-else block
    \item a loop region (LR) which is composed of a loop block
    \item a sequential region (SR) which is composed of any two consecutive regions
    \item a basic block region (BBR), which can have one or more lines of sequential code
\end{itemize}
As regions are hierarchical, they are composed of other regions. For example, an if-else block inside a for loop is a branch region inside a loop region. The outermost region in the hierarchy represents the entire program or a code fragment of interest.

\begin{figure}
\centering
\includegraphics[width=3in,height=1in]{figures/region_nos.jpg}
\caption{Region representation of a program}
\label{fig:region}
\end{figure}

Figure ~\ref{fig:region} shows a code fragment with its region representation. At the top, the whole program is a sequential region composed of one basic block region and a loop region. The loop region in turn is composed of a basic block region and a branch region. The branch region is further composed of a true region (the if part) and a false region (the else part). Both of these are basic block regions.

\textbf{}
\textbf{fix indentation in Figure 2, and add line numbers for easy reference, and add the line numbers in description above}
}

\subsection{Compile Time Optimizations}
\label{comopt}

Pandas programs can be optimized by manually rewriting them using appropriate API features. However,  many users are not aware of these features. The availability of multiple frameworks and differences in their APIs further complicates the problem. Our framework performs a variety of optimizations during compile time, i.e.\ the static analysis and rewrite phase. 

Optimization transformations can be applied only if certain preconditions are met.  Static analysis provides information about the preconditions, as well as other information needed to carry out the transformations.

Once the source is transformed to \SCIRPy, our system can utilize static analysis tool-sets provided by Soot. For example, Soot's control flow graph (CFG) module is used to construct the CFG of Python programs. Live variable analysis (LVA) \cite{muchnik1997}, which is provided by Soot, can be used  to find which variables are live at any point in the program.
A variable is \textit{live at a program point} if there exists a path from that point to the exit of the program along which the current value of the variable may be used.  We use a modified version of live variable analysis, called \textit{live attribute analysis} (\textit{LAA}), which checks liveness of individual attributes (columns) of dataframes. Even if a dataframe is live at a program point, some of its columns may not be used subsequently and are thus not live.   The result of the analysis is used to perform column selection optimization. Details are discussed in Section ~\ref{opt}.


\fullversion{
Compatibility with Soot also allows us to integrate existing code of DBridge ~\cite{c12} project \textit{for example for live attribute analysis??}.
}


\begin{figure}

\begin{lstlisting}[language=Python]
import lazyfatpandas.pandas as pd
pd.analyze() # transfer control to LaFP
df = pd.read_csv('data.csv', parse_dates=['tpep_pickup_datetime']) #fetch data 
df = df[df.fare_amount > 0] # filter bad rows
# add features
df['day'] = df.tpep_pickup_datetime.dt.dayofweek
# aggregation
df = df.groupby(['day'])['passenger_count'].sum() 
print(df) # use dataframe
\end{lstlisting}
\caption{Sample Program}
\label{l:p1}
\end{figure}

\begin{figure}
\vspace*{3mm}
\begin{lstlisting}[language=Python]
import lazyfatpandas.pandas as pd
from lazyfatpandas.func import print # LaFP's lazyPrint
SO_columns = ['pickup_datetime','passenger_count','fare_amount']
df = pd.read_csv('data.csv',usecols=SO_columns)
df = df[df.fare_amount > 0]
df['day'] = df.pickup_datetime.dt.dayofweek
df = df.groupby(['day'])['passenger_count'].sum()
print(df)
pd.flush() # calls compute()
\end{lstlisting}
\caption{Optimized Version of Program in Figure ~\ref{l:p1}}
\label{l:op1}
\end{figure}


To understand the benefits of LAA, consider the Pandas program in Figure ~\ref{l:p1} (from ~\cite{Magpie}), which fetches data from files into in-memory dataframes.  Then different computations like data filtering, feature addition, and aggregation are performed on it. 
The optimized version (output of static analysis phase) of the program is as shown in Figure ~\ref{l:op1}. 

The original program fetches all 22 columns from the test dataset. Only 3 of these are used in the program, which is inferred by live attribute analysis, as discussed in Section ~\ref{opt:DS}.  The optimized program after applying the column selection optimization fetches only the required 3 columns, by passing them in the usecols option to read\_csv. 




\begin{figure}

\begin{algorithmic}
\Function{pd.analyze}{()}
\State source\_file $\leftarrow$ get\_source\_code()
\State \SCIRPy\ $\leftarrow$ python\_to\_\SCIRPy(source\_file)
\State opt\_\SCIRPy\ $\leftarrow $ static\_analysis\_opt(\SCIRPy) 
\State \hspace*{5mm} /* Performs Static optimization + code rewrite to
\State \hspace*{1cm} support runtime optimization */
\State opt\_code $\leftarrow $ \SCIRPy\_to\_python\_opt(opt\_\SCIRPy)
\State executor (opt\_code)
\EndFunction
\end{algorithmic}
\caption{Just-in-Time Static Analysis}
\label{jit}
\end{figure}

\subsection{Just-in-Time Static Analysis}
\label{ssec:jit_static}

One of the novel contributions of our approach is the Just-in-Time (JIT) static analysis, which performs static analysis at the start of program execution.
The static analysis and optimization is initiated by calling the analyze() method of our lazyfatpandas library as the very first step of the program.

Our JIT approach provides an easy and convenient way for users to perform static-analysis based optimizations. Other static analysis tools require users to perform static analysis and program rewrite as a separate phase, following which the rewritten program must be executed.  In contrast, our approach does not require any change in the flow of code optimization/execution.

The process of JIT static analysis is described in Figure~\ref{jit}.  The pd.analyze() method is called before execution of user code.  
The analyze() method identifies the source program code, parses it, converts it to \SCIRPy, and performs static analysis and compile time optimizations.  
Further, code transformations are performed during this phase to enable run time/lazy optimizations discussed in Section ~\ref{opt}.  The
optimized IR is converted back to Python and executed.



\subsection{Lazy Evaluation and Task Graphs}
\label{lazyscalableruntime}
  
In eager evaluation, an expression is evaluated as soon as it is reached during program execution. In contrast, in a lazy evaluation strategy, when an expression is reached during program execution, instead of evaluating it, an expression node is created, and added to a task graph.  The task graph is evaluated only when it is needed, by calling a function that forces the evaluation to occur.  Spark and Dask, as well as our Lazy Fat Pandas (LaFP) framework, are examples of systems based on lazy evaluation of dataframes; evaluation is forced only when a function such as  \texttt{compute()} is called to actually execute the operations.

\begin{figure}
\centerline{\includegraphics[bb=0 0 270 437.04, width=.29\textwidth]{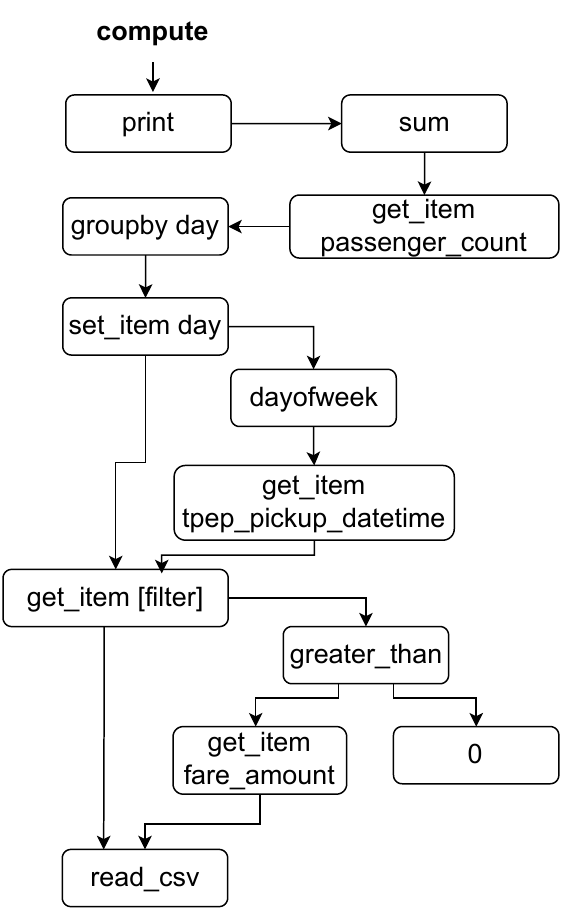}}
\caption{LaFP Task graph of the Program in Figure ~\ref{l:p1}}
\label{fig:task_graph_eval_img}
\end{figure}

A task graph is a directed acyclic graph (DAG) in which the nodes represent a computational task or operation and the edges denote the precedence constraints among these computational tasks.
The task graph for the program in Figure ~\ref{l:p1} is shown in Figure ~\ref{fig:task_graph_eval_img}. 
An edge \textbf{($A \rightarrow B$)} represents that the task $B$ 
depends on task $A$.  Such an edge may be created when the result of the operation at node B is an input for the operation at node A, and is also used to enforce output order for lazy print statements.\footnote{The direction of the edge follows the convention for task graphs and dependency graphs, although the flow of data is in the opposite direction.}

The optimized source code has calls to lazy versions of the Pandas API calls, defined in our Lazy Fat Pandas (LaFP) framework, which supports a (subset of) the Pandas DataFrame API, but using the LaFPDataFrame.  A call to the LaFP API function does not immediately execute the operation.
Instead, each operation creates a new FatDataFrame object, which is then linked to the task graph based on its inputs.  The lowest level operation is generally a read data operation which represents a lazy DataFrame created lazily from the input dataset.   

The task graph is executed only when an operation result needs to be passed to an operation that needs a materialized (non-lazy) data frame, or at the end of the program, when computation can no longer be deferred.


A key motivation for lazy evaluation is that task graphs can be transformed to optimize evaluation, at runtime, just before being executed; most lazy frameworks support such optimization. 
For example, redundant nodes (expressions) can be pruned to avoid computations that are not used, and nodes can be re-arranged for early data filtering (similar to predicate pushdown) to improve performance.  Optimizations implemented in LaFP are discussed later in Sections~\ref{ssec:runopt} and ~\ref{opt}.





Once run-time optimizations are performed in LaFP, we have an optimized task graph, which is then executed using any of the backends supported by LaFP including Pandas itself, Dask and Modin.
LaFP allows the backend to be chosen in the program.

For small datasets that fit in memory, Pandas is usually faster than the lazy frameworks and is the preferred choice.  In this case LaFP optimizations help speed up the program compared to directly running it on Pandas.
If the backend chosen is lazy, it may perform its own optimizations; in that case the optimizations performed by LaFP complement the backend optimizations.  
(We are currently working on automating the choice of backend based on memory usage estimates.)

One of the distinctive features of our approach compared to just 
using a lazy framework is that it is  capable of utilizing information generated using JIT static analysis phase. This information provides look-ahead beyond points where lazy computation cannot be deferred further, allowing us to detect
for example that some columns are not used later in the program and therefore can be projected away, or not even computed earlier in the execution.  
Static-analysis based rewriting also allows our framework to detect which API calls can handle LaFP data frames; all other 
API calls default to Pandas dataframes, and our optimizer introduces calls to force computation before invoking such APIs.


\fullversion{
Task graph nodes record \textcolor{red}{information such as source nodes, operation performed by the node, } which columns are used from the inputs, as well as columns output by the operation; these are used for
avoiding creation of columns that are not needed, or dropping of columns that are no longer needed.
}

\fullversion{
Note that the task graph may be disconnected either because of independent operations, or because of operations that need to be evaluated to allow API calls that cannot consume lazy dataframes, where the computed dataframe may also be used subsequently as inputs to other lazy dataframe operations.
}

\fullversion{
Each node in the task graph maintains the following information:
\begin{enumerate}

    \item \textbf{Sources:} List of nodes that are the input of this operator
    \item \textbf{Action:} Type of action of the operator along with all arguments passed
    \item \textbf{AllAttributes:}  List of all DataFrame attributes (columns) with their data types
    \item \textbf{usedAttributes:} Attributes that are used in the node
    \item \textbf{modAttributes:} Attributes that are modified in the node
    \item \textbf{Usecols:} List of all the attributes used in the computation till the node in task graph
    \item \textbf{Useful:} Whether the node affects the final computation used by the user
\end{enumerate}
}



\subsection{Run-Time Optimizations and Execution}
\label{ssec:runopt}

LaFP allows the task graph to be executed on any one of the supported backends, which are currently Pandas, Dask, and Modin, with the default being Dask.
The user can select the required back-end by just adding one line of code, for example:
 \newline
pd.BACKEND\_ENGINE=pd.BackendEngines.PANDAS

The choice of the backend significantly impacts execution time and memory usage for a program. Pandas and Modin use an eager evaluation approach with all data required to be in memory (or distributed memory in the case of Modin).  In contrast, Dask employs lazy evaluation, and supports data sizes larger than memory.

Support for multiple back-ends for the same program provides the users with an opportunity to select the best framework applicable for their program, depending on factors such as the size and type of datasets, available hardware resources etc. The optimal back-end can also be identified in a cost-based manner, implementation of which is a part of future work.

Traditional databases perform several optimizations when a query is submitted and execute it based on an efficient query plan performing only the necessary computation. Lazy frameworks such as Spark and Dask also optimize the task DAG before execution.  Examples of optimizations include removal of unused computations, pushing selections and projections, and
fusing of operators to reduce data movement.


Similarly, the runtime module of LaFP performs several optimizations in the task graph (DAG) at run time.  
Optimizations include Predicate Push Down, Redundant operations elimination, Lazy Print etc that offer significant performance benefits.
Metadata is also used, when available, to reduce storage costs by making use of more space efficient data types. 
Computation of metadata is done as a background task.

LaFP also offers the option to use any one of the supported back-end executors. The optimized task graph is executed using the selected executor.
When a lazy backend such as Dask is chosen, the optimizations performed in LaFP complement optimizations in the lazy backend.  When an eager backend such as Pandas itself or Modin is chosen, the backend cannot perform optimization across nodes, and thus LaFP optimizations are even more important.


The execution of the task graph in LaFP is done as follows,
for the case where the backend is eager.
The LaFP task graph is executed in topological order. After evaluating a task graph node, the result is stored in a field called \texttt{result}. This \texttt{result} field is cleared once all its dependent nodes have been evaluated, and the result is no longer needed, minimizing memory usage. This is managed by counting the in-degree of each task graph node before executing the task graph and decrementing the count after a node is used to generate another node's result. When the count reaches zero, the \texttt{result} field is cleared, allowing Python's garbage collector to reclaim the memory. As the task graph is executed from bottom to top, the results of lower nodes, which have been evaluated and used, are deleted to keep memory usage to a minimum.


For example, consider the task graph in Figure \ref{fig:task_graph_eval_img}, \texttt{read\_csv} is executed first, and the result is stored in the node. This result is cleared after its dependent nodes \texttt{get\_item fare\_amount}, and \texttt{get\_item [filter]}, have been executed. Finally, the \texttt{result} of the node on which \texttt{compute} was called is returned. If \texttt{compute}() is called on a print node, then \texttt{None} is returned.

In case the backend is a lazy framework such as Dask, instead of executing the operation when traversing the task graph,
the API call is transformed to the compatible API call for the selected lazy backend.
The execution of the operations on the lazy backend is initiated when root of the task graph is reached,
or when the results are required for an intermediate operation such as print. 
Nodes whose results are used more than once can be persisted using persist() on dask dataframe,
avoiding recomputation for subsequent uses.

For backends other than Pandas, LaFP performs some transformations to deal with incompatibilities between Pandas and the selected backend.
For example, pandas 'read\_csv' API call supports a keyword argument 'index\_col' to specify the column(s) to use as row labels for the Dataframe. 
Dask Dataframe does not support this keyword argument. However, similar behavior can be obtained by making another API call 'set\_index' on the Dask Dataframe after 'read\_csv'. 
Our framework is capable of identifying such inconsistencies across multiple frameworks. 
Our framework performs this additional operation in Dask transparently such that the end result is the same after the execution of 'read\_csv' API calls irrespective of the back-end.



\section{Static and Runtime Optimizations}
\label{opt}

In this section, we discuss a number of static and run time optimizations implemented in LaFP. These optimizations perform holistic optimization of the program. These optimizations are performed in two settings. 
\begin{itemize}
    \item By rewriting of imperative programs based on static analysis
    \item By optimizing the generated task graph at runtime, before it is executed.
\end{itemize}
Some of the optimizations we describe exploit a combination of information from static analysis and
runtime information.



\subsection{Column Selection}
\label{opt:DS}

In many programs, not all the columns (attributes) from the input dataset are used. Fetching such unused columns into the memory leads to increased memory usage and extra IO operations. Column selection optimization identifies and fetches into memory only those columns that are used later in the program. Live attribute analysis (LAA), discussed below, helps to identify which columns are used.  Once LAA is performed, the relevant API calls are modified to fetch only these used columns into memory.  
 
\textit{Live attribute analysis (LAA)}
is an extension of the well known technique of live variable analysis, as mentioned earlier in Section ~\ref{comopt}. LAA treats attributes (columns) of a dataframe as a variable. Similar to a variable, a dataframe column is live at a program point if there is a path to the program exit along which it may be used. However, columns are different from variables in that 
assignments or uses can happen at the level of entire dataframes: 
\begin{enumerate}
    \item If the whole dataframe is used at a program point, all columns of that dataframe become live.
    \item Similarly, all columns of a dataframe are killed at the point of definition of a dataframe.
    \item If a dataframe is derived from another dataframe, its liveness information is used to determine liveness information for the source dataframe.
\end{enumerate}

Live variable analysis is done by using dataflow analysis, which is based on the Gen and Kill sets at each node in the control flow graph (CFG).
The dataflow equations for live attribute analysis, which are modified from those for live variable analysis, are as below:
\begin{equation}
\begin{split}
Gen_n = \{ d.i \mid & i \textrm{ is a column of dataframe } d, \textrm{ and either } \\
& d.i \textrm{ or all of } d \textrm{ (without specifying any  } \\
& \textrm{ column) is used in basic block } n,  \textrm{ prior to } \\
& \textrm{ any assignment to } d.i \textrm{ or to } d \}. 
\end{split}
\end{equation}
\begin{equation}
\begin{split}
Kill_n = \{ d.i \mid & i \textrm{ is a column of dataframe } d, \textrm{ and either } \\
& d.i \textrm{ or all of } d \textrm{ (without specifying any ) } \\
& \textrm{ column is assigned in } n \}
\end{split}
\end{equation}

\eat{

\begin{equation}
  Gen\textsubscript{n} = \left\{
                \begin{array}{ll}
                  d.i\| \textrm{ i column of dataframe d is used in basic} \\
                        \textrm{block n and that use is not preceded  by a }  \\
                        \textrm{definition of dataframe d in the basic block n.}\\
                  
                  \textrm{}\\
                  d.\tau\|\textrm{dataframe d is used fully in the basic block}\\
                          \textrm{n and that use is not preceded by a definition} \\
                           \textrm{of dataframe d in the basic block n.}
                \end{array}
              \right\}
\end{equation}

\begin{equation}
      Kill\textsubscript{n} = \left\{
                \begin{array}{ll}
                  
                  \textrm{d.}\tau \|\textrm{dataframe d is defined in basic block n} \\
                  \textrm{}\\
                  \textrm{d.i} \| \textrm{i attribute of dataframe d is defined in }\\
                  \textrm{basic block n } \\
                  
                \end{array}
              \right\}
\end{equation}
}


Note that if a dataframe is passed as an attribute of a function called from $n$, we assume that all columns of the dataframe are used in $n$.  Global variables pose another challenge, and if a dataframe is assigned to a global variable, we assume conservatively that all its columns are used in any function called from $n$.  Further, aggregate operations kill all
columns except those used in the aggregate or in the groupby operation.

We next define sets $In_n$ and $Out_n$ which merge local information provided by $Gen_n$ and $Kill_n$, with information from successor nodes of $n$, to identify global liveness information.
\begin{equation}
Out\textsubscript{n} =   \bigcup_{s \in succ(n)} \textit{In}\textsubscript{s}
\end{equation}
\begin{equation}
In\textsubscript{n} =  Gen\textsubscript{n} \cup  \{ \textit{Out} \textsubscript{n}- \textit{Kill} \textsubscript{n}  \} 
\end{equation}
\par
The above equations are solved to get the Gen, Kill, In and Out sets for each basic block $n$.   The live attributes at the end of a
basic block $n$ are those that are in the set $Out_n$. 
$In_n$ represents liveness information immediately before the block and $Out_n$ represents liveness information immediately after the block.

Once LAA is performed, liveness information is available for all columns of all dataframes at all program points.  The column selection optimization modifies the IR to fetch only those dataframe columns that are live (in Out$_n$) of the program point $n$ where the dataframe is created from an input dataset, e.g. by a read\_csv() call.
\fullversion{
The column selection algorithm is shown in Algorithm~\ref{algo:data_select}.
}

We now consider how live attribute analysis works on the program in Figure ~\ref{l:p1}. This program has only one  dataframe, i.e., df. 
The last statement of the  program prints the dataframe, so all columns are live at 'In' of this point.
Line 8 results in only columns day and passenger\_count being live.
At line 6, the column pickup\_datetime 
becomes live, whereas column day is killed as it is assigned and thus not alive before that.  Line 4 makes fare\_amount live.
The columns live at 'Out' of the Line 3 are 'pick\-up\_datetime', 'passenger\_count' and 'fare\_amount', and only these need to be read from
the csv file.
The optimized version of the program, which reads only the above columns, is shown in Figure ~\ref{l:op1}. 

We also note that informative API functions df.head(), df.info() and df.describe() are frequently used to get an idea 
of the dataset contents and, their output does not affect the intended program result.  Treating these
as using all attributes of df would result in unnecessary column retrieval, so as a heuristic we ignore the
attribute usage of these functions.


\fullversion{
\begin{algorithm}[bt!]
\caption{Column Selection Algorithm} 
\label{algo:data_select}
\begin{algorithmic}
\Procedure{\textbf{Column\_Selection()}}{}

\State \SCIRPy\  $\leftarrow$ Transform program to \SCIRPy
\State cfg $\leftarrow$ buildCFG(\SCIRPy) 
\State perform\_LAA()
\ForAll{$ unit \in cfg$}
    \If{ unit $=$ df creation stmt}
        \State dfAttrs $\leftarrow$ getDFLiveAttrs(unit)
        \LineComment Create a list stmt with these attrs 
        \State stmt $\leftarrow$ getListStmt(dfAttrs, 'SO\_columns')
        \State insert\_stmt\_bef(stmt,unit)
        \LineComment Update unit to fetch these attributes
        \State update\_df\_stmt(stmt, unit, 'usecols')
   \EndIf 
\EndFor
               
\EndProcedure
\end{algorithmic}
\end{algorithm}
}



\subsection{Predicate Push Down}
\label{chapter:row_selection}

Performing selection operation early in databases, also known as predicate push down, reduces the size of relations and therefore reduces the computation to be performed during other operations like joins. In dataframe systems, filter operations reduce the size of datafame. We identify the filter operations in the task graph, and move them as close to the data source as possible. 


Predicate pushdown in dataframe systems is more complex than in traditional database systems since the operators are more complex, and further, task graphs are DAGs, not trees.

We therefore define safe points, and push filter operations as early as possible, but only beyond these safe points, to ensure that the task graph's semantics are not modified.


We define \textit{safe point(s)} for an operation as point(s) beyond which it can be moved without potentially modifying the output of the program.
Let $mod\_attrs(u)$ be columns that $u$ either modifies or computes, and let $used\_attrs(u)$ denote the attributes that are used by the operator $u$ (including use
by UDFs invoked by $u$).

Given a dataframe operator $u$ and a filter operation $f$ which 
uses the output of $u$, we can swap $u$ and $f$ in the task graph, under the following conditions: 
\begin{enumerate}
    \item $mod\_attrs(u) \cap  used\_attrs(f) = \phi$ 
    \item filtering the rows in the input to $u$ does not affect the computed values for the output rows of $u$ that satisfy filter $f$
    \item $f$ is the only parent of $u$.
\end{enumerate}
for a number of operators, including getitem, setitem, sort\_values, drop\_duplicates, filter, rename, replace, type conversion operators 
such as to\_datetime, to\_frame, astype or convert\_dtypes, fillna, round, abs among others.




We note that there are operators like explode, and merge can affect the number of rows, and we do not allow predicates to be pushed below such operators.  Predicates cannot be pushed through operators whose semantics are not known, and thus for which we cannot compute mod\_attrs or used\_attrs.
Further, we cannot push predicates below operators that
have side effects, for example operators that perform
output operations.

We also note that in the special case where $u$ has multiple
parents, if all the parents have the same filter $f$, and
the other conditions for pushing the predicate below $u$ are satisfied, we can push $f$ below $u$, and remove the filter
operator $f$ from all the parent paths.
On the other hand, if all parents of $u$ have filter operators,
with predicates say $p1, p2, \ldots, pn$ respectively,
we can push a predicate $p1 \wedge p2 \wedge \ldots \wedge pn$
below $u$, while retaining the filters above $u$.

Finally we note that while our predicate pushdown is done on the task graph, it is also possible to infer predicates and push them 
through control flow graphs by using static analysis and 
rewriting programs, similar in spirit to the predicate pushdown performed by MagicPush \cite{MagicPush}.
This can enable some cases of predicate pushdown beyond
points where computation is forced.  However, it is not
clear how much benefit this will provide beyond predicate pushdown on task graphs enabled by lazy evaluation, and we have not implemented it currently.




\fullversion{
\begin{algorithm}[bt!]
\caption{Predicate Pushdown Algorithm}
\label{algo:row_selection}
\begin{algorithmic}

\Procedure{\textbf{PredicatePushdown}}{task\_graph \textbf{G}}
  \ForAll {node \textit{v} in post-dfs order of task graph G}
    
    \LineComment {Push down the filter to earliest point possible}
    
    \State $u \gets $ Immediate preceding operation of v in task graph

    \LineComment {u preceeds v in task graph (program start...u-v...end)}
     \If { (v is a row\_selection statement) \\
        \quad\quad\quad\quad \textbf{and} (u is not a row\_selection statement) \\
        \quad\quad\quad\quad \textbf{and} (has\_swap\_conflict(u,v) is \textbf{False}) \\
        \quad\quad\quad\quad\textbf{and} (v is live at all program paths from u to end)}
        
       \State Remove v from all other paths from u to end
       \State Swap(u,v) and \textbf{continue}
    \EndIf
    \EndFor
\EndProcedure
\end{algorithmic}
\end{algorithm}

\begin{algorithm}[bt!]
\caption{Check swap conflicts for Predicate Pushdown}
\label{algo:has_swap_conflicts}
\begin{algorithmic}
\Procedure{\textbf{has\_swap\_conflicts}}{task\_graph \textbf{u},task\_graph \textbf{v}}
  \ForAll {node \textit{v} in post-dfs order of task graph G}
  
    \LineComment {u preceeds v in task graph (program start...u-v...end)}
    \If {(used\_attributes(v) $\cap$ mod\_attributes(u)  $\neq \phi)$ \\
    \quad\quad\quad\quad\textbf{OR} (operator(u) in [Operators affecting df shape]) \\
    \quad\quad\quad\quad\textbf{OR} (operator(u) in [Agg, Op requires complete pass of data])}
    \State \Return True
    \EndIf
    \State \Return False
    \EndFor
\EndProcedure
\end{algorithmic}
\end{algorithm}
}

\subsection{Lazy Print}\label{sec:lazy_printing}

Dataframe computations in lazy frameworks are deferred until computation is forced by a call to a \texttt{compute()} (or similar) method. 
Computation needs to be forced when passing dataframes to functions, for example print, that cannot accept lazy data frames.
If compute can be deferred, the task graph could
includes later parts of the code, which can enable
other optimizations that 
may not be possible if compute has to be done earlier.
Thus, postponing invocation of \texttt{compute()} can help to reduce the execution cost.

Print is one of the most common functions that forces computation.  One approach to delay prints, which we tried initially is to move print statements to later in the program based on static analysis, coupled with the use of a temporary variable to hold the data needed for printing. However, this method is relatively complicated and can be expensive or fail if the dataframe to be printed is very large.

Instead, we introduce a lazy version of the print operator, allowing compute calls to be deferred beyond the (lazy) print calls.  With our novel approach, lazy print statements are treated as operations and added to the task graph.
When the task graph is executed, the print nodes are processed, and the data is printed.
However, care must be taken to ensure that outputs are
generated in the correct order, and we now describe how we enforce that.

We note that although we have currently only implemented the lazy version of print, the approach is quite general, and can be used for example for generating plots, and for other output operations.

\begin{figure}
\begin{lstlisting}[language=Python]
import lazyfatpandas.pandas as pd
pd.analyze()
df = pd.read_csv("data.csv")
print(df.head())  # lazy print
df["day"] = df.pickup_datetime.dt.dayofweek
p_per_day = df.groupby(["day"])["passenger_count"].sum()
print (p_per_day) # lazy print
avg_fare = df.fare_amount.mean()
print(f"Average fare: {avg_fare}")
\end{lstlisting}
\caption{Program with Multiple Print Statements} 
\label{lazy_print_ex}
\end{figure}

\begin{figure}
\begin{lstlisting}[language=Python]
import ...
from lazyfatpandas.func import print # override print with lazy print
...[optimized read csv as in Figure 4}
print(df.head())  # lazy print
df['day'] = df.pickup_datetime.dt.dayofweek
p_per_day = df.groupby(['day'])['passenger_count'].sum()
print(p_per_day)  # lazy print
avg_fare = df.fare_amount.mean()
print(f'Average fare: {avg_fare}')  # lazy print
pd.flush() 
\end{lstlisting}
\caption{Optimized Program with Lazy Print} \label{lazy_print_ex_o}
\end{figure}

\begin{figure}
\vspace*{5mm}
\includegraphics[width=0.33\textwidth]{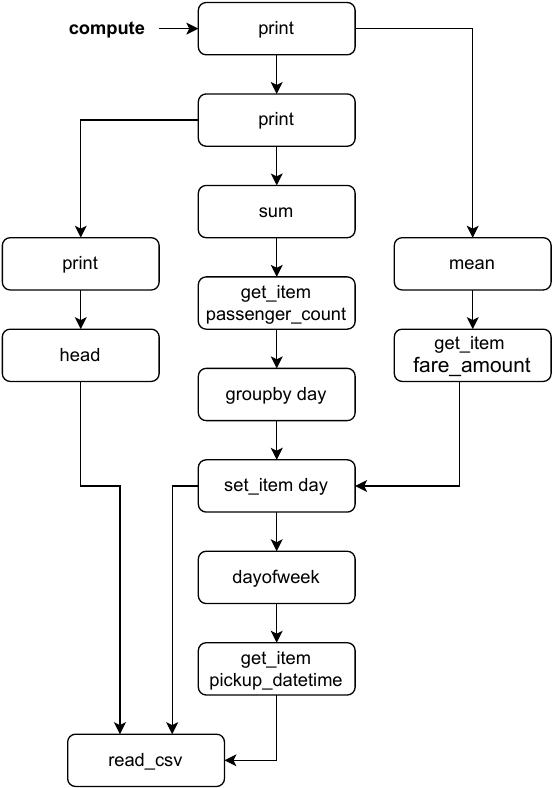}
\caption{Task Graph for Program in Figure \ref{lazy_print_ex_o}}
\label{fig:lazy_print_ex_img}
\end{figure}

Figure~\ref{lazy_print_ex} presents an example with multiple print statements. Its optimized version is shown in Figure~\ref{lazy_print_ex_o} and the equivalent task graph is shown in Figure~\ref{fig:lazy_print_ex_img}.  The optimized program overrides the built-in print method with LaFP's lazy print method 
by importing print from lazyfatpandas.func.
The library lazyfatpandas.func also provides lazy versions of some other functions; for example the lazy version of Python's len()
function, when applied to a lazy dataframe, returns a lazy integer, else it behaves like the normal len() function.

\fullversion{
We can also use from lazyfatpandas.func import *, which imports lazy versions of print as well as other functions like len()
}

When LaFP's lazy print is called, the node representing the print operation is added to the task graph, with the lazy dataframes as the source nodes. A dependency edge is added to the previous print operation (if any) to maintain the correct print order. All the lazy dataframes used in the print statement are identified and appropriate edges added to the task graph to ensure that these dataframes are computed before the lazy print is (eventually) executed. 


At the end of the program, \texttt{pd.flush} is called, which internally invokes \texttt{compute} on the last print node, forcing the computation of the task graph. The print statements are processed in the correct order due to the dependency edges between print nodes. The statements to override print, as well as 
the call to \texttt{pd.flush()} are automatically inserted in the source program by program rewriting, thus fully automating the process.


Python allows objects, including dataframes, to be used in Python's formatted strings (f-strings), for example:\\
\hspace*{1cm}\verb|print(f'Average fare: {avg_fare}')| \\
in Figure~\ref{lazy_print_ex}, where \verb|avg_fare| is a dataframe.   Creation of the formatted string would require the dataframe to be computed.  

To defer the computation of the formatted string, while retaining the link to the correct dataframe (since the variable may get assigned in a subsequent step before the print is executed) the "lazyprint()" wrapper function replaces the dataframe variable by the unique ID of the task graph node representing the dataframe, along with an escape sequence to mark the unique ID. 

When the constructed string is processed by the execution of the deferred print function, the function checks for the escape sequence to identify the unique ID of the task graph node.  Further, this node must be computed before the lazy print is processed at the end of the program, which is ensured by the runtime. 
\fullversion{
For example, in line 10 of Figure~\ref{lazy_print_ex}, the dataframe \texttt{avg\_fare} is used in an f-string. The string that will be constructed is \texttt{"Average fare: \$\_\#UID\$\_\#"}. When lazy prints are processed, \texttt{"\$\_\#UID\$\_\#"} is replaced by the value of the computed dataframe.
}


\subsection{Forced Computation for External Module Function Invocation}
\label{sec:force_compute}

When a program in a lazy framework calls a function that expect an evaluated Pandas dataframe, computation needs to be forced before the dataframe is passed to such a function call. An example of such a commonly used function is \texttt{matplotlib}.  Programmers using frameworks such as Dask have to manually insert code to force computation.  

To deal with the above issue, as part of program rewriting, a compute() call is added to force computation before execution of any function call for which a lazy implementation is not available; the materialized (computed) dataframe is then passed to the function call.

\begin{figure}
\begin{lstlisting}[language=Python]
import lazyfatpandas.pandas as pd
#external module
import matplotlib.pyplot as plt
pd.analyze()
df = pd.read_csv("data.csv")
print(df.head())  # lazy print
df["day"] = df.pickup_datetime.dt.dayofweek
p_per_day = df.groupby(["day"])["passenger_count"].sum()
print (p_per_day) # lazy print
plt.plot(p_per_day)
plt.savefig("fig.png")
avg_fare = df.fare_amount.mean()
print(f"Average fare: {avg_fare}")
\end{lstlisting}
\caption{Program with External Module Invocation} 
\label{force_compute_ex}
\end{figure}

\begin{figure}
\begin{lstlisting}[language=Python]
import ...
from lazyfatpandas.func import print # lazy print
...[optimized read csv as in Figure 4]
print(df.head())
df['day'] = df.pickup_datetime.dt.dayofweek
p_per_day = df.groupby(['day'])['passenger_count'].sum()
print(p_per_day)
#pending lazy prints evaluated and 
#p_per_day is computed and passed to plt.plot
plt.plot(p_per_day.compute(live_df=[df]))
plt.savefig('fig.png')
avg_fare = df.fare_amount.mean()
print(f'Average fare: {avg_fare}')
#evaluate remaining lazy prints 
pd.flush()
\end{lstlisting}
\caption{Optimized Version of Program From Figure~\ref{force_compute_ex}}
\label{force_compute_ex_o}
\end{figure}

\fullversion{
\begin{figure*}
  \centering

\begin{subfigure}{0.33\textwidth}
\includegraphics[width=\textwidth,height=10cm]{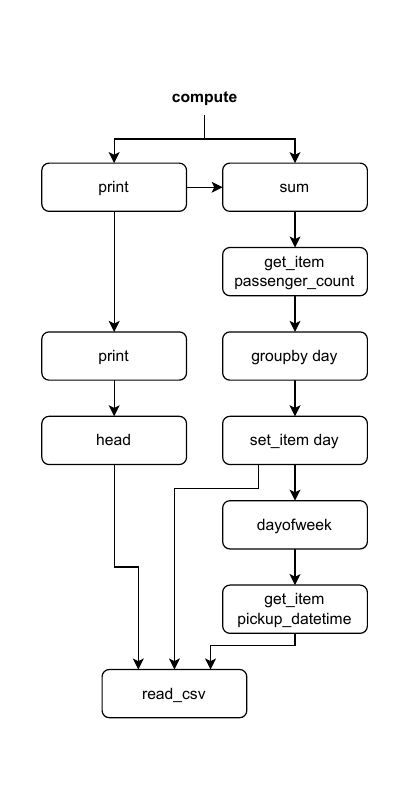}
  \subcaption{Forced Computation in line 13 of Figure \protect\ref{force_compute_ex_o}}
  \label{fig:force_compute_img1}
\end{subfigure}
\begin{subfigure}{0.33\textwidth}
\includegraphics[width=\textwidth,height=10cm]{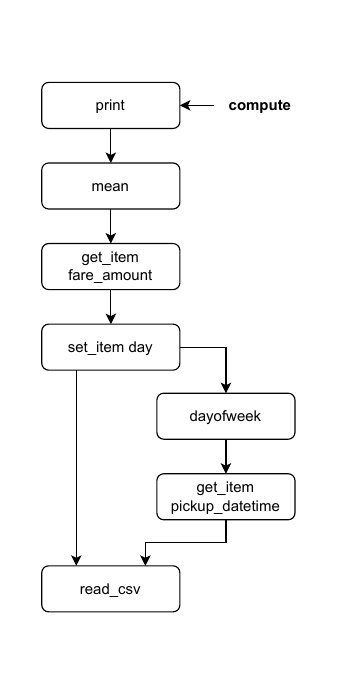}
  \subcaption{Remaining Task Graph of Figure \protect\ref{force_compute_ex_o}}
  \label{fig:force_compute_img2}
\end{subfigure}
\caption{Task Graphs for Lazy Print and Forced Computations}
\label{fig:AllTaskGraph}
\end{figure*}
}


The programs in Figure~\ref{force_compute_ex} and  its optimized version in Figure~\ref{force_compute_ex_o} demonstrates forced computation.  Our static rewriting optimizer keeps track of all the external modules imported during program rewriting, for example \texttt{import matplotlib.pyplot as plt}.  The optimizer checks whether a dataframe is passed as an argument when a function from any such modules is invoked. If so, a call to \texttt{compute} is added; for example line 10 of Figure~\ref{force_compute_ex} is rewritten as shown in  Figure~\ref{force_compute_ex_o}. (The passed argument live\_df=[df] is used in another optimization which is discussed later).

\fullversion{
\begin{figure}
\begin{lstlisting}[language=Python]
# df is the source dataframe
df1 = pd.read_csv(data.csv) 
# operations on a dataframe yield a new dataframe
df2 = df.head()             
# operations on a dataframe yield a new dataframe
df3 = df[df.column > 0]     
# return type is a dataframe
df4 = func(...)             
\end{lstlisting}
\caption{Cases Where New Dataframe Variables are Created}
\label{force_compute_ex1}

\end{figure}
}

To invoke \texttt{compute} on a dataframe, we need to figure out which variables are dataframe variables.
This information is inferred from the types of the Pandas API calls.
For example, Pandas functions like read\_csv() or read\_parquet(), as well as most Pandas functions 
on dataframes return dataframes.

\fullversion{
To infer the dataframe variables, we use following rules:
\begin{enumerate}
    \item Source dataframe is created using \texttt{read\_csv}, \\ \texttt{read\_parquet} or similar APIs.
    \item Any operation on a dataframe except in-place operation returns a dataframe or a Series. 
\end{enumerate}
These cases are illustrated in Figure~\ref{force_compute_ex1}.
Our rewrite system uses these rules to keep track of dataframes that are created or updated in the program.
}

Further, since external modules like \texttt{matplotlib.pyplot} can generate output, which can conflict with lazy printing since the output order may get changed. 
To solve this issue, pending print operations are processed when a dataframe is forced to compute, maintaining the correct output order.




In line 10 of Figure \ref{force_compute_ex_o}, when \texttt{p\_per\_day.compute()} is invoked, the task graph containing all
the lazy calls up to that point, including lazy prints, are executed, 
and the result of the node \texttt{sum} (\texttt{p\_per\_day}) passed to \texttt{plt.plot}, 
generating a plot image. 
\fullversion{
The corresponding task graph shown in Figure \ref{fig:force_compute_img1}. 
}
Subsequently, when \texttt{pd.flush} is called in line 15, the print in line 13 is processed, along with lazy operations that
are pending.
\fullversion{
The task graph is shown in Figure ~\ref{fig:force_compute_img2}.
It may be noted that the sub-tree rooted at node \texttt{set\_item day} is recomputed when \texttt{pd.flush} is called.
}
Note that the shared subexpression corresponding to the dataframe df computed in line 5 would get recomputed on further execution in the program when pd.flush() is called.
When the shared subexpression is first computed, information related to the future usage of the dataframe in 
the rest of the program is unavailable to the lazy framework.
We can use static analysis to avoid recomputation, as discussed next in Section (~\ref{al:ccr}).


\subsection{Common Computation Reuse}
\label{al:ccr}



As discussed in the previous section, forced computation  can lead to re-computation of shared sub-expressions.
We can avoid re-computation by persisting dataframes that are used in more than one place, before and after a force computation
boundary.  However, in lazy evaluation frameworks, information about future reuse beyond the current 
task graph is not available when the computation is forced so we do not know which dataframes will be reused. 
Naively persisting every intermediate result just in 
case it is reused is not practical since it would drastically increase memory footprints.

Therefore, we make use of static analysis to identify useful (live) sub-expressions to be cached when compute is invoked on a dataframe. 
We perform Live DataFrame Analysis (LDA) which is similar to Live Attribute Analysis (discussed earlier in Section ~\ref{opt:DS}) to identify live (useful) dataframes at each program point.  When we force computation, we check which dataframes are live after the point when the computation is forced, and provide the list of such dataframes to the compute method, 
which persists (caches) any common sub-expressions between the expressions defining the live dataframes and the nodes in the task graph being computed.   

When the task graph is executed, any node marked for persistence
has its result persisted on its first execution.
In later executions, the persisted result is reused, instead of being
recomputed.

In Figure ~\ref{force_compute_ex_o}, the compute method is called on p\_per\_day, and an argument named live\_df is introduced, and static analysis is used to generate the  list of dataframes live after that program point and pass it as the value for the live\_df argument in line 10.   
After line 10, df is the only live dataframe, and it is used later to compute avg\_fare. 
Since df is a shared sub-expression (common node) between both p\_per\_day and avg\_fare, the compute call
at line 10 includes the parameter ``live\_df = [df]'', which is live and a common subexpression, so it will be cached during computation.

\fullversion{
Note that caching is done on the task graph nodes.  
Consider the task graphs in Figure~\ref{fig:lazy_print_ex_img}.
The forcing of computation of p\_per\_day requires part of the
task graph 
in Figure~\ref{fig:force_compute_img1} to be computed.
This leaves a residual task graph shown in Figure~\ref{fig:force_compute_img2} to be computed later.
The node labeled set\_item\_day corresponds to the dataframe df
identified earlier as live, and is common to the part of the 
task graph computed earlier, and the part computed later.
This node is marked for persistence when p\_per\_day is computed. 
}


Once all uses of a persisted dataframe have been completed, it can be safely discarded to release memory. Our lazy computation framework discards persisted dataframes after their last use when they are no longer subexpressions of dataframes in live\_df list passed to the compute(). 

\subsection{Using and Computing Metadata}
\label{opt:MA}


Data type information and data statistics are both very important for efficient computation.  
While relational databases provide strong typing, and precompute statistics such as number of rows, average row size, as well as more detailed statistics like number of distinct values in each column, or histograms on columns, such information is not available for file-based data formats.  The widely used csv format does not even provide type information; formats such as Parquet provide type information, but not statistics.

We compute metadata for each source data file.   Statistics can be computed from a sample of the values in a collection. To get correct datatypes we have to scan the entire file, although we can do it based on the first few rows or a sample, at some risk.  
The metadata for a file is computed by running a script on the file, and stored for later use.
Information like  modified time,  column names and types, approximate size of each row, and approximate number of rows in the dataset are currently maintained in the metadata.

The metadata for a file is accessed when a dataset is used in a program. If the meta-data about the dataset is available in the metastore, it is retrieved, and used.  The modified time metadata is used
to ensure that if a file has been updated, metadata computed before the last update is not used, since it may be outdated.

Decisions on what framework to use depend on whether the dataframes can fit in memory, which can be inferred from the metadata statistics such as number of rows, sizes of rows, data types, etc.   We are currently working on automated choice of backend based on comparing required space for dataframes (based on metadata statistics) with available free space.

Pandas infers some datatypes such as integers and floats automatically (defaulting to int64 or float64) when reading data from csv files, to avoid the less efficient string representation.   
If the programmer knows the type, it can be specified using the dtype attribute of read\_csv() function, which can reduce storage costs significantly in some cases.
Our read\_csv() wrapper uses the type information from metadata, and passes it to the read\_csv() function of the backend, using the dtype argument.

One of the commonly used type-related optimizations is to replace string types by category types, when there are only a small number of possible string values.   
A collection of values is created, and the strings are 
represented by (small) integer values indexing into the collection. 
If the type of a column is specified as category,
read\_csv() creates a category type containing the required values.  
From the  metadata, we can infer if a column has a small number of values and is a candidate for category type.  

However, if we assign datatype as category, and the program subsequently assigns a new value not in the input dataset to one of the rows in the dataframe, there would be a run-time error.  

To ensure that the category type is used only when safe to do so, we make use of the kill information from static analysis (Section~\ref{opt:DS}) to check which columns of each dataframe are read-only, i.e. they are never assigned after being read.   The list of read-only columns can be
passed to our read\_csv() wrapper function, which then makes use of metadata and declares a column a  category type if it has only a small number of distinct values, and is read-only.  (The read-only check is currently under implementation.)

One of the differences between Pandas as compared to Dask/Spark is that the latter need datatypes in many cases.  For example, the df.apply() function in Dask requires the output data type to be specified using the meta parameter, if the system is not able to infer it.  We can use the metadata to statically infer types, and pass them to the relevant Dask functions.  Type inferencing is an area of ongoing work.

\section{Related Work}
\label{sec:relwork}


Dask \cite{rocklin2015dask} provides abstractions over NumPy Arrays, Pandas Data\-frames, and regular lists, allowing dataframes to be partitioned, and operations on dataframes to run in parallel. Dask operations can work on partitions of data, with dynamic and memory-aware task scheduling to achieve parallel and out-of-core execution. Modin ~\cite{tsds} provides a "drop-in" replacement for Pandas, and supports parallel/multi-core execution. 
PySpark \cite{pysparkpandas} supports the 
Pandas DataFrame API on top of Apache Spark.  
Ray \cite{ray} is another framework for distributed programs, which supports scalable datasets (dataframes) among other features. Modin uses Ray as one of its back-end options.

Multiple libraries use lazy evaluation to optimize data science applications.  Cunctator ~\cite{cun} delays API function evaluations to identify opportunities for API fusion for optimization. Similarly, DelayRepay ~\cite{delayrelay} optimizes GPU specific workloads by using lazy evaluations over NumPy APIs. Weld ~\cite{weld} is another tool that supports cross-platform optimizations by providing an IR to represent workloads of different libraries and by performing optimizations across libraries. To the best extent of our knowledge, none of these support optimizations discussed and implemented by us. 


Like \cite{self_paper} we perform program rewriting based on static analysis using Soot.  However, they only consider compile-time optimization, without JIT static analysis and do not consider lazy evaluation or other run-time optimizations, which are among our key contributions.  Live attribute analysis is mentioned in \cite{self_paper} but without any details.


Dias~\cite{DIAS} accelerates Python notebook based programs by replacing code sequences with semantically equivalent but faster code sequences. It uses a pattern matcher and a rewriter. The Dias rewriter applies rewrite rules dynamically at run time, one cell of a notebook at a time, and injects checks to verify the correctness of rewrites. The \SCIRPy\ optimizer works on a similar principle, however, in comparison to Dias where cell-by-cell code transformations are performed, \SCIRPy\ analyzes the entire program for optimization opportunities. In addition, LaFP provides a lazy run time for task graph based optimizations. LaFP also supports multiple back-end for transparent execution of the same program over large dataset.

Magicpush ~\cite{MagicPush} uses a generate-and-verify based approach for predicate pushdown in data science applications.  It uses symbolic execution and verification and can handle programs having non-relational operators and user-defined functions (UDFs) for wider applicability. In contrast, LaFP performs predicate pushdown optimization by identifying safe points in the programs. Also, LaFP performs a wide variety of other optimizations.

\fullversion{
The DBridge project~\cite{batchopt2008,dbridge2011,batchasync2015} focused on optimization of database applications written in Java by rewriting of application code.  Rewriting was done by transformation rules based on static analysis.
Transformation of imperative code to SQL was addressed in \cite{imptosql2017} (using static analysis) and \cite{qbs2013} (using query synthesis).

Optimizations implemented include batching of queries to reduce network roundtrips \cite{batchopt2008}, asynchronous submission to avoid waiting on round trip delays \cite{async2011,holistic2012,batchasync2015}.
Subsequent work on the DBridge project focused on translation of computation carried out by imperative code on data fetched from a database to database queries in SQL \cite{equivsql2016,imptosql2017,cobra2018}. 
Similarly, Query Based Synthesis(QBS) ~\cite{qbs2013} transforms imperative code fragments into SQL queries. \textsc{Casper} ~\cite{caspermapreduce2018} uses the same technique to transforms imperative Java code into MapReduce paradigm.
}

\fullversion{
LegoBase ~\cite{legobase}, optimizes programs by performing source-to-source compilation of high-level Scala code to low-level C code. 
Tuplex ~\cite{tuplex} is a framework that compiles Python UDFs into optimized native code with fall-back option for exceptional cases. 
All these libraries scale out the dataframes by making use of multiple cores and lazy evaluation. To the best of our knowledge, none of these focus on identifying  the inefficiencies introduced in the program by the way it is written. For example if a program is using some of the columns of a dataset but fetching all of them in the dataframe, then LaFP transformation techniques will eliminate fetching the unused columns, which will not be the case in other platforms.
}

\section{Performance Evaluation}
\label{sec:perf}

\eat{
\begin{figure*}

  \centering
  \begin{subfigure}{0.48\textwidth}
\includegraphics[width=\textwidth,height=4cm]{figures/benchmark_figl/lm0.pdf}
  \subcaption{Memory Consumption (450 MB Dataset)}
  \label{fig:MaxMemory1}
\end{subfigure}
\begin{subfigure}{0.48\textwidth}
\includegraphics[width=\textwidth,height=4cm]{figures/benchmark_figl/lt0.pdf}
  \subcaption{Execution Time (450 MB Dataset)}
  \label{fig:time1}
\end{subfigure}
\begin{subfigure}{0.48\textwidth}
\includegraphics[width=\textwidth,height=4cm]{figures/benchmark_figl/lm1.pdf}
  \subcaption{Memory Consumption (1.4 GB Dataset)}
  \label{fig:MaxMemory1}
\end{subfigure}
\begin{subfigure}{0.48\textwidth}
\includegraphics[width=\textwidth,height=4cm]{figures/benchmark_figl/lt1.pdf}
  \subcaption{Execution Time (1.4 GB Dataset)}
  \label{fig:time1}
\end{subfigure}

\begin{subfigure}{0.48\textwidth}
\includegraphics[width=\textwidth,height=4cm]{figures/benchmark_figl/lm2.pdf}
  \subcaption{Memory Consumption (4.2 GB Dataset)}
  \label{fig:MaxMemory2}
\end{subfigure}
\begin{subfigure}{0.48\textwidth}
\includegraphics[width=\textwidth,height=4cm]{figures/benchmark_figl/lt2.pdf}
  \subcaption{Execution Time (4.2 GB Dataset)}
  \label{fig:time2}
\end{subfigure}

\begin{subfigure}{0.48\textwidth}
\includegraphics[width=\textwidth,height=4cm]{figures/benchmark_figl/lm3.pdf}
  \subcaption{Memory Consumption (12.6 GB Dataset)}
  \label{fig:MaxMemory3}
\end{subfigure}
\begin{subfigure}{0.48\textwidth}
\includegraphics[width=\textwidth,height=4cm]{figures/benchmark_figl/lt3.pdf}
  \subcaption{Execution Time (12.6 GB Dataset)}
  \label{fig:time3}
\end{subfigure}
\caption{Maximum Memory And Time Consumption for different programs for various Datasets on different platforms}
\label{fig:AllBenchmark}
\end{figure*}
}



In this section, we study the benefits of our optimization techniques on a variety of programs, across different dataset sizes.

\subsection{Benchmark Programs and Datasets}

To benchmark LaFP, we have taken real workloads from a variety of sources including programs used in ~\cite{DIAS}~\cite{Magpie}~\cite{MagicPush} . These programs execute a variety of operations like data filtering, data augmentation using feature addition, data aggregation, data merge, etc., analyzing data from domains such as movie rating systems, taxi data, startup analysis  etc.\footnote{We have shared the programs and datasets anonymously at https://anonymous.4open.science/r/LaFP\_Benchmark-B1FB, and will release our benchmark publicly later.}

Each program was executed on datasets of size 1.4 GB, 4.2 GB, and 12.6 GB to study the impact
of dataset size.  We replicated or pruned the original datasets that were available with the programs, 
to create datasets of approximately the target size (in most cases within 5\% of the target size).

For each program, we compare the performance with and without our optimizations (including both run time and rewrite optimizations), using Pandas, Dask and Modin as the back-ends. 
For comparison with direct use of Dask and Modin frameworks, we manually transformed the programs to use Dask and Modin.
Executing Pandas programs on Modin is straightforward, with the only change required being to an import statement.
To get Pandas programs to run on Dask, the programs usually needs to be rewritten, for reasons such as the need for
forcing computation in a lazy framework, lack of support for some API methods, and more fundamentally, the fact that 
Dask does not preserve ordering of rows in the dataframe, and thus position-based indexed access does not work.
To check the performance with Dask, we modified the programs manually to make them compatible with Dask. 
Some programs required minor modifications like passing appropriate datatypes and code changes to force compute before certain API calls (like sort\_index()), 
others required rewriting to work around APIs, such as iloc() for row position, of features such as inplace updates, that are not supported by Dask dataframes. 

Modin can use different execution engines, such as Ray or Dask. We use Ray as the default executor for Modin and LaFP Modin programs. In some cases where a program/dataset combination could not be executed using Ray, we use Dask as executor for Modin. For such programs, Dask is used as executor for LaFP Modin as well for uniform comparison.

We present end-to-end performance results for LaFP;  
it is difficult to segregate out benefits of compile time versus run time optimizations
since LaFP utilizes static and runtime optimizations in a unified manner. 
Optimizations like common computation reuse, forced computation, lazy print utilize both run time and compile time 
steps.  

We run all our experiments on a hexa-core AMD Ryzen 5 3600 with base clock at 3.6 GHz with 32GB of DDR 4 3200Mhz RAM.


\begin{figure}
\centering
{\small  
\begin{tabular}{|c|c|c|c|c|c|c|} 
 \hline
 Size & Pandas & LPandas & Modin & LModin & Dask & LDask\\ [0.5ex] 
 \hline
 1.4GB & 10 & 10 & 10 & 10 & 10 & 10\\ 
 \hline
 4.2GB & 10 & 10 & 9 & 9 & 10 & 10\\ 
 \hline
 12.6GB & 2 & 7 & 4 & 7 & 8 & 9\\ 
 \hline
\end{tabular}
}
\caption{Number of Programs Successfully Executed on Different Platforms}
\label{tab:nos_program_executed}
\end{figure}

\subsection{Applicability of LaFP}
\label{sec:perf:applic}

A major goal of our optimizations was to ensure that Pandas programs can run successfully even on large datasets.  
The first set of experiments therefore checks how many programs could complete successfully using the different
frameworks, with and without optimization.

Our framework allows any backend to be used without any program rewrite (barring an import of the lazyfatpandas library, 
a configuration line to choose the backend, and a call to analyze()). 
The changes required to enable execution using Dask are implemented by a combination of rewriting based on static analysis 
(for example to force computation), and via wrapper functions in the LaFP API.
The wrapper functions uses appropriate Dask calls to implement Pandas API functionality where possible, and in other cases 
(such as inplace updates, column rename, shape changing operations etc)
converts dataframes to Pandas, applies the function in Pandas, and converts the dataframe back to Dask.
Passing input/output types to apply() function is required in Dask, and adding it automatically is under implementation;  
we manually added the type for one program that needed it.

Users however need to be aware that using Dask may affect the order of rows in dataframes, which may
potentially affect subsequent operations that are order sensitive.
If the program has such order sensitive operations, the user should not choose Dask as the 
backend.\footnote{Automating the detection of order sensitivity is an area of future work.}

Figure~\ref{tab:nos_program_executed} shows how many programs/dataset combinations could execute successfully on different backends,
with LPandas, LModin and LDask denoting the versions using our rewrite+runtime optimizations, using, respectively, Pandas, Modin and Dask as backend executors.

\begin{figure}
\centering
\includegraphics[width=3.5in,height=2in]{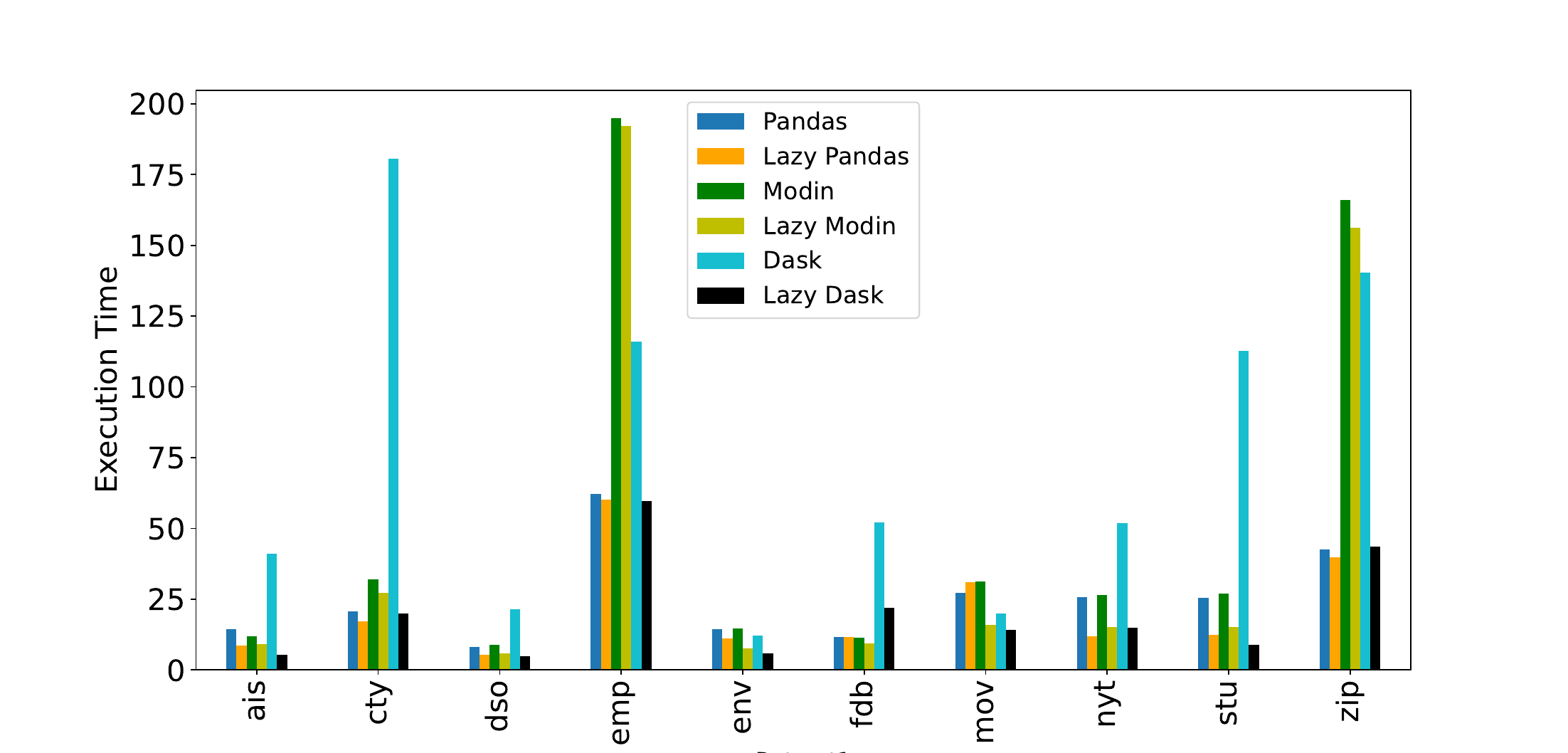}
\caption{Execution Time on Different Platforms - 1.4 GB}
\label{fig:executiontimrall}
\end{figure}

\fullversion{
\begin{figure}
\centering
\includegraphics[width=3.5in,height=2in]{figures/benchmark_fig_n/executiontime_450.pdf}
\caption{Execution Time on Different Platforms - 450 MB}
\label{fig:executiontimrall}
\end{figure}

\begin{figure}
\centering
\includegraphics[width=3.5in,height=2in]{figures/benchmark_fig_n/executiontime_150.pdf}
\caption{Execution Time on Different Platforms - 150 MB}
\label{fig:executiontimrall}
\end{figure}
}


With Pandas, only 2 programs run successfully  with 12.6 GB dataset. Modin could execute only 4 such programs. Dask could execute 8 programs. The Pandas/LPandas and Modin/LModin programs that failed had run out of memory.  The one program where Dask/LDask also failed was the 'emp' program; on inspection we found that there was a call to an external plot function which required materializing a large dataframe as a Pandas dataframe, which resulted in out of memory error.
In contrast, using our optimizations, all except the above program, i.e. 9 programs out of 10, ran successfully with Dask as backend (i.e. LDask). Only 7 programs ran with either Pandas or Modin as backend (i.e., LPandas or LModin).  The improvements in the case of Pandas and Modin were because our rewrites could reduce space usage, whereas un-optimized Pandas/ Modin programs ran out of memory. 

We also built a regression test framework to ensure that the datasets computed with our optimizations were identical to the results on
Pandas without any optimization, by computing and comparing hashes (computed using md5) of the dataset results; our optimized programs on different platforms
all passed these tests.
\fullversion{
\footnote{Interestingly,  the default Python hash function gives different results on each run. a known behavior, so we use the md5 function
as the hash function.}
}


\setlength{\belowcaptionskip}{0pt}
\setlength{\abovecaptionskip}{-1pt}

\begin{figure}
\begin{subfigure}{0.48\textwidth}
\includegraphics[width=\textwidth,height=4cm]{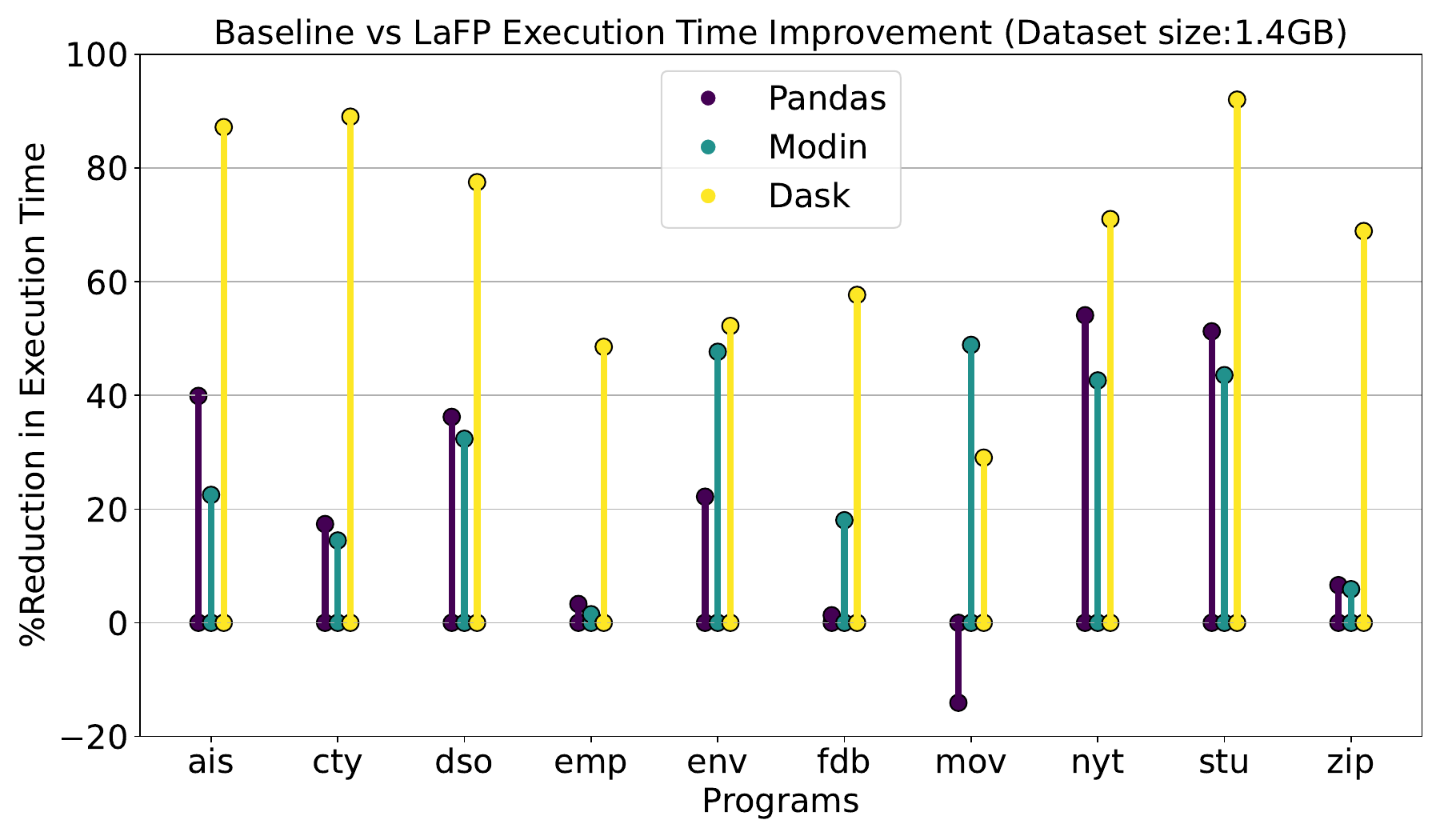}
  \subcaption{1.4 GB Dataset}
  \label{fig:time1}
\end{subfigure}
\begin{subfigure}{0.48\textwidth}
\includegraphics[width=\textwidth,height=4cm]{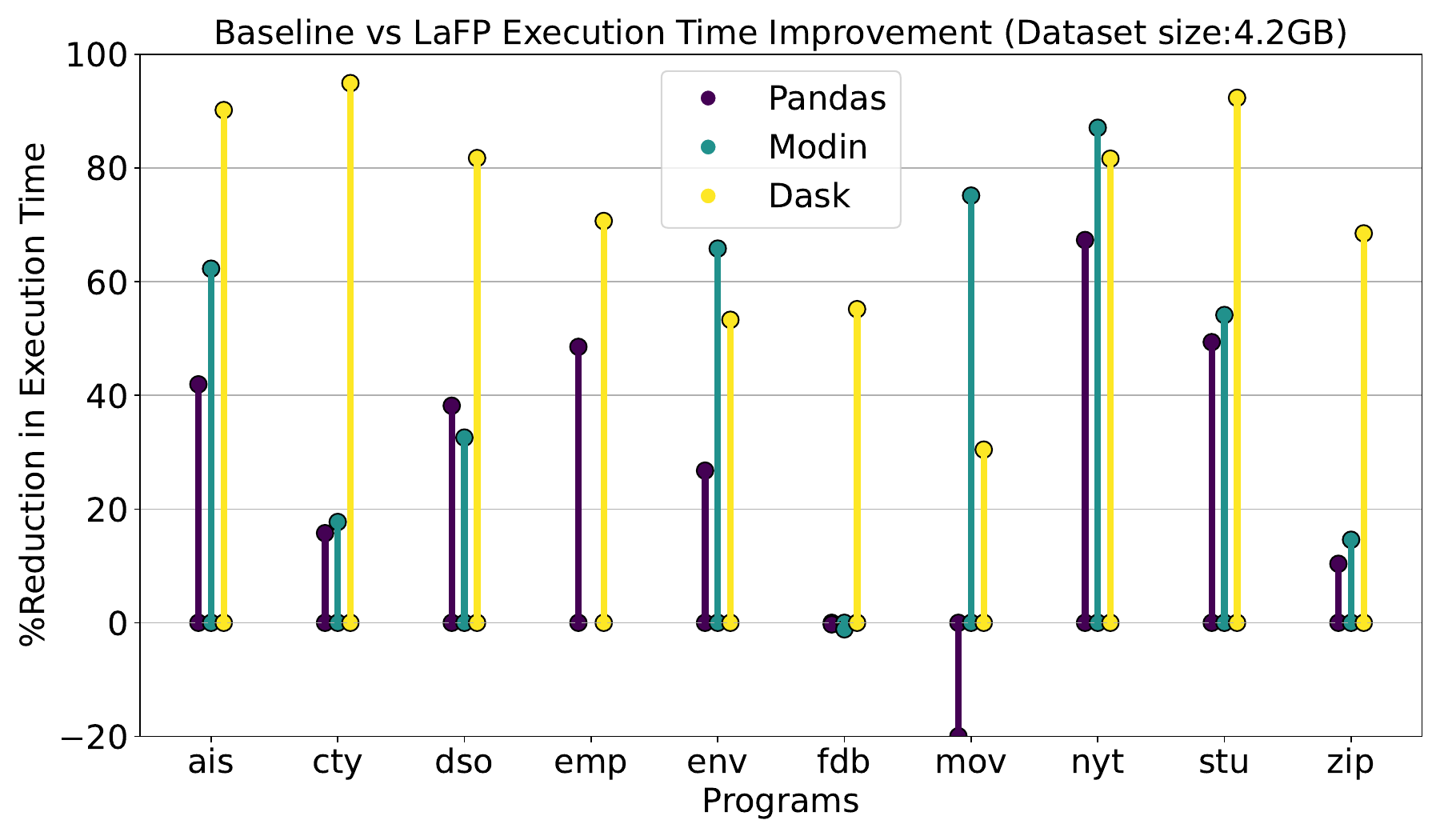}
  \subcaption{4.2 GB Dataset}
  \label{fig:time2}
\end{subfigure}
\begin{subfigure}{0.48\textwidth}
\includegraphics[width=\textwidth,height=4cm]{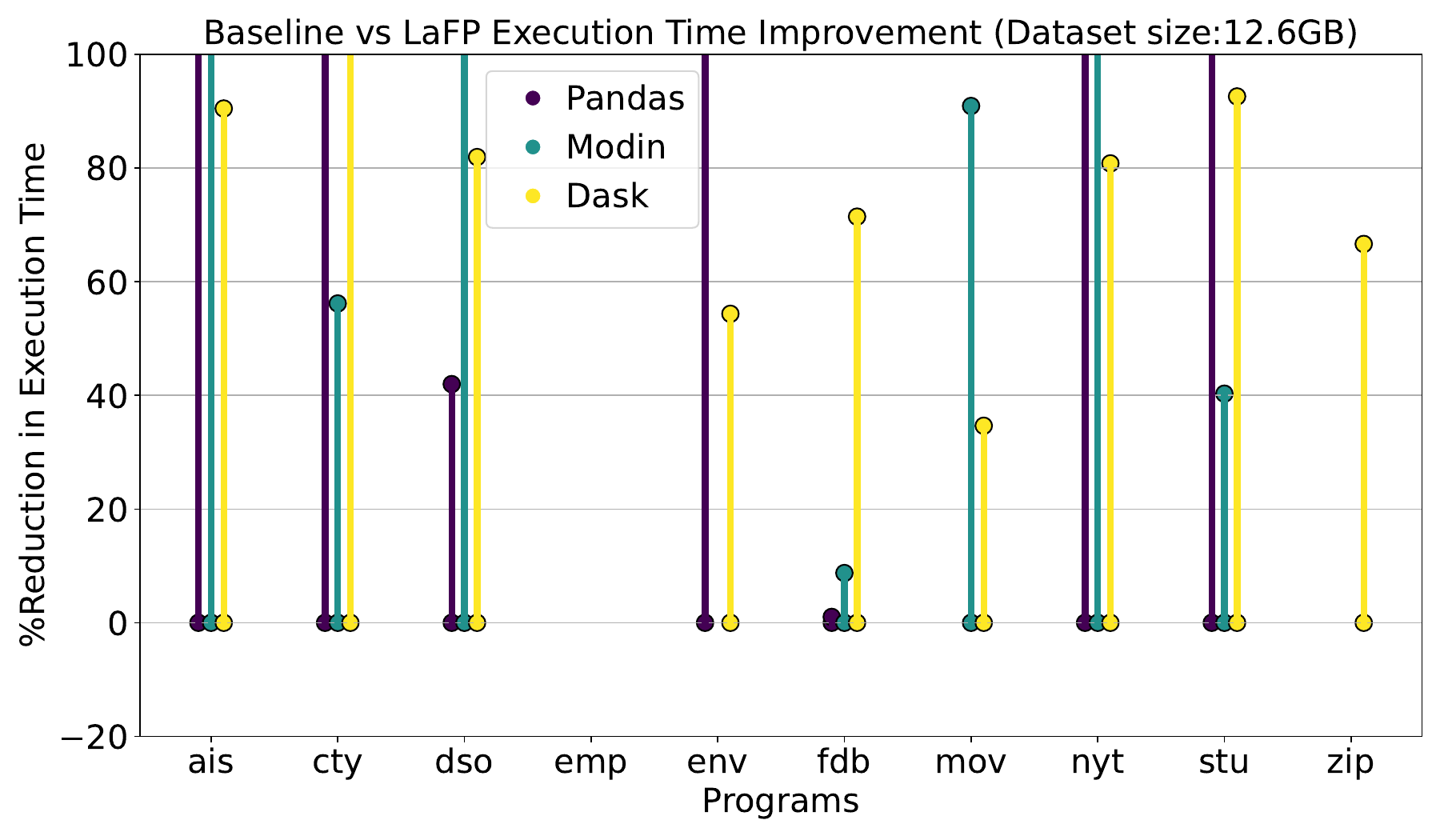}
  \subcaption{12.6 GB Dataset}
  \label{fig:time3}
\end{subfigure}
\caption{Execution Time Improvement}
\label{fig:perf:time}
\end{figure}

\begin{figure}
\begin{subfigure}{0.48\textwidth}
\includegraphics[width=\textwidth,height=4cm]{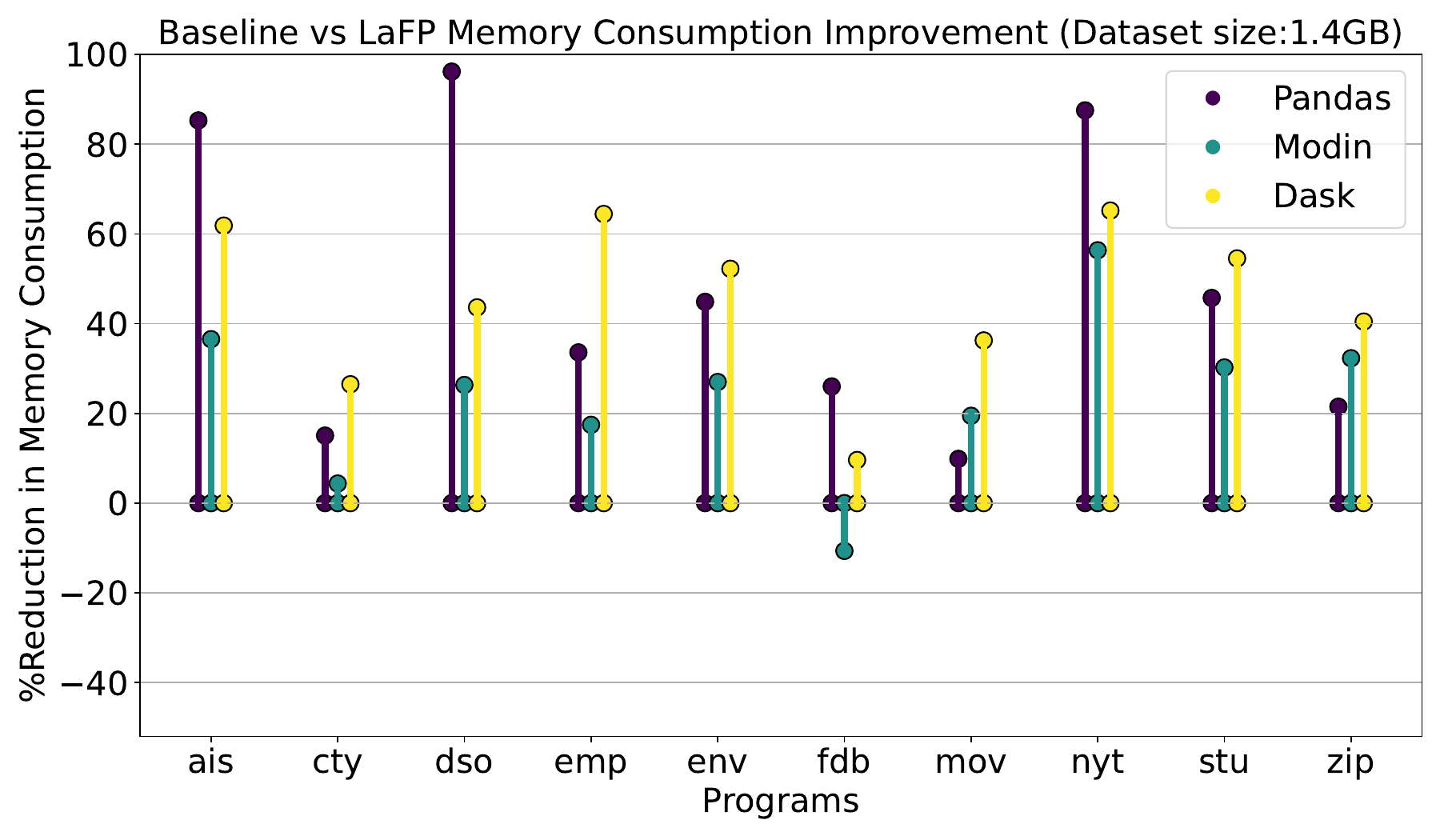}
  \subcaption{1.4 GB Dataset}
  \label{fig:MaxMemory1}
\end{subfigure}
\begin{subfigure}{0.48\textwidth}
\includegraphics[width=\textwidth,height=4cm]{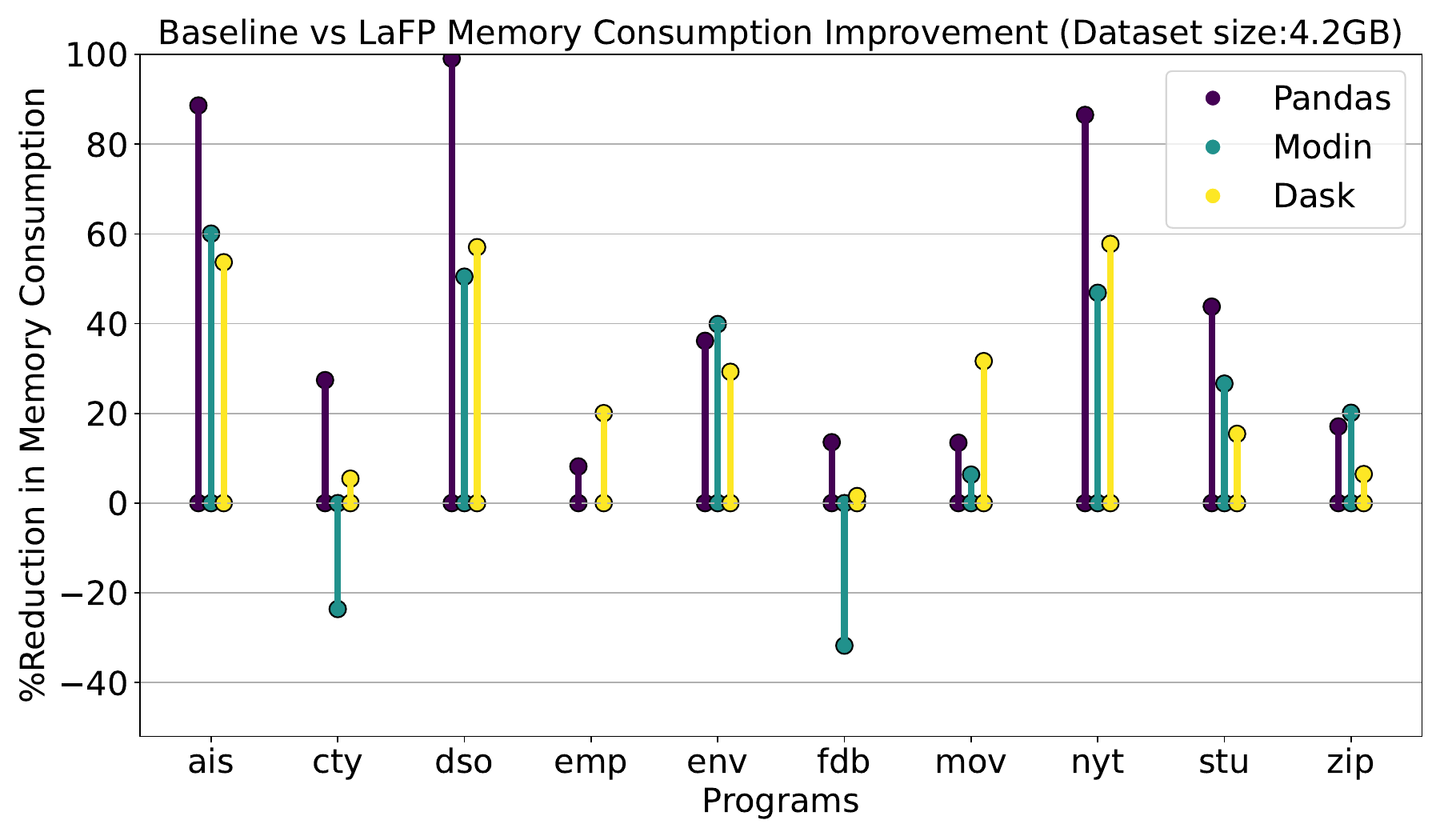}
  \subcaption{4.2 GB Dataset}
  \label{fig:MaxMemory2}
\end{subfigure}
\begin{subfigure}{0.48\textwidth}
\includegraphics[width=\textwidth,height=4cm]{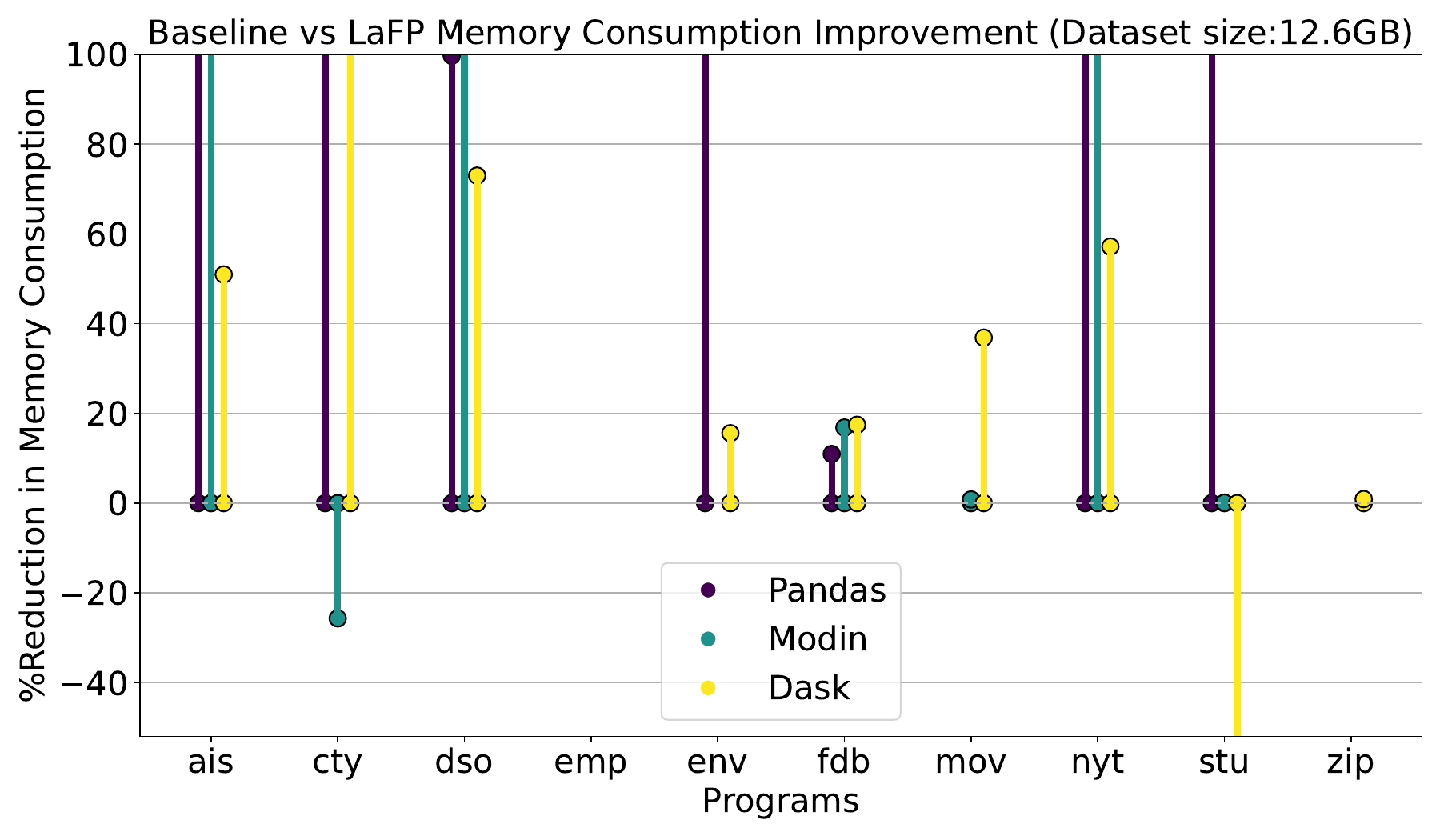}
  \subcaption{12.6 GB Dataset}
  \label{fig:MaxMemory3}
\end{subfigure}
\caption{Memory Consumption Reduction}
\label{fig:perf:mem}
\end{figure}

\subsection{Execution Time}

We first consider absolute execution times for different configurations, on the 1.4 GB dataset, where all the configurations could run successfully. 
The results are shown in Figure~\ref{fig:executiontimrall}.
It can be seen that Pandas and Modin are faster than Dask when datasets fit in memory.  The optimized versions of the programs running on LaFP are better than the original versions in almost all cases.  Dask is slower than Pandas in most of the cases, but interestingly, Lazy Dask versions are not only faster than Dask, 
but also faster than all the alternatives in most of the cases, presumably because of the combination of our optimizations with Dask runtime optimizations.
However, for smaller datasets (whose results are not shown due to lack of space), Lazy Pandas was faster than all other alternatives in many of the cases.

Next, we consider larger datasets, where Pandas and Modin are not able to run successfully in several cases.
Figure~\ref{fig:perf:time} shows the execution time improvements using our optimizations, as a percentage of the original time, across different backends (Pandas, Modin and Dask), for  different programs for different dataset sizes.

In cases where the corresponding program and dataset combination could not execute successfully on the relevant backend, we treat the original execution time
as infinity, resulting in 100\% improvement in 
performance. 
Correspondingly, missing data points in the Figure~\ref{fig:perf:mem}, such as for 'emp' on 12.6GB, represent cases where the program could be not be executed using the relevant backend, nor could the
optimized version of the program be executed with the same backend.
It can be seen that across Pandas, Modin, and Dask, there were multiple programs that, without our optimizations, could not be executed on the 12.6 GB dataset. 

On the datasets where the backends worked successfully, our optimizations resulted in significant improvement in execution time on almost all cases, with upto nearly 70\% reduction (3X speedup) in execution time on Pandas, and upto 90\% improvement (10X speedup) with Modin, and upto 95\% improvement (19X speedup) with Dask.
There were very few cases where our optimizations increased the execution time, with the worst case being 20\% more time as compared to the original Pandas program (shown as a negative 20\% improvement).

Although Dask has its own run-time optimizer, our optimized versions running on Dask performed significantly better in almost all cases.
Column selection rewrite played a significant part in improving the performance.
We also found that for applications that have multiple prints, our optimizations provided particularly high improvements on Dask, 
thanks to our innovative lazy print optimization.

Multiple optimizations are applicable on several of the benchmarks. For example, on 'stu' program, metadata analysis, Lazy print and common computation reuse via 
caching are applicable.  
Using Dask backend, with 12.6 GB of data, the LaFP optimizations reduced the execution time by 13X.  Turning off caching reduced the time benefit to 1.4X,
demonstrating the significant benefit of this optimization.



We also measured the overhead due to our static analysis and rewriting optimization techniques.
The time taken by JIT static analysis phase and 
rewriting for various programs is in the range of 0.04 sec - 0.59 sec, which is a very small fraction of the execution times of the programs.

\subsection{Maximum Memory Consumption}
\label{Per:MaxMem}

To get reliable memory size estimates despite multi-threaded execution with Modin and Dask, we created a separate thread
that monitored the process memory usage every 200 milliseconds, and we report the maximum memory usage.
Figure~\ref{fig:perf:mem}  shows the memory consumption improvements using our optimizations, as a percentage of the original memory usage, across different backends (Pandas, Modin and Dask), for  different programs for different dataset sizes.

In cases where the corresponding program and dataset combination could not execute successfully on the relevant backend, we treat the memory consumption
as infinity, resulting in 100\% improvement in 
performance.  Missing values in the Figure~\ref{fig:perf:mem} represent cases where neither the original nor the optimized program 
could execute successfully.

On cases where Pandas could run successfully, our optimizations improve memory consumption significantly in most cases, with over 
95\% reduction in some cases where column selection was particularly helpful. 
For the programs that can be executed with Modin, our optimizations reduced memory consumption by upto 60\%. 
There were only a few cases where our optimizations increased memory consumption, with 25\% increase in the worst case.



LaFP with Dask reduces memory utilization compared to Dask in many cases, with upto 
up to 70\% reduction in memory consumption, primarily due to projection pushdown rewriting.
However, in other cases LaFP has increased memory utilization;  for example with 'stu' programs on 12.6 GB of data persisting
of common subexpressions resulted in 2.3X increase in memory consumption, while providing 13X speedup in execution time.
If we disable common sub-expression caching the memory consumption reduces to 80\% of that of the baseline, but the execution 
speedup drops from from 13X to 1.4X.
It may be noted that persisted dataframes are memory resident not only when using Pandas and Modin, 
but also with Dask when using Dask's persist() API function, which results in increased memory usage.
Persisting Dask dataframes on disk is an area of future work.

\section{Conclusion and Future Work}
\label{sec:concl}

We have described novel optimization techniques for Pandas programs, that combine the best of two approaches: static analysis based rewriting and lazy runtime frameworks.  Our performance study shows the significant benefits of our optimization approach in terms of its ability to successfully execute Pandas programs on large datasets, in terms of execution speeds, and in terms of memory usage, as well as its superiority to just using lazy frameworks.


Future work includes implementation of automated choice of backend based on size estimates and order dependency, completion of read-only attribute analysis 
to ensure category type can be safely used, type inference for input/output parameters to be passed to Dask apply() function, and addition of support for
more Pandas API calls.
Since Modin and Dask both support clusters of machines, execution on a cluster of machines and related performance evaluation is part of our future work.

 \vspace{20pt}
 



\bibliographystyle{ACM-Reference-Format}
\bibliography{sample}


\begin{thebibliography}{17}


\ifx \showCODEN    \undefined \def \showCODEN     #1{\unskip}     \fi
\ifx \showDOI      \undefined \def \showDOI       #1{#1}\fi
\ifx \showISBNx    \undefined \def \showISBNx     #1{\unskip}     \fi
\ifx \showISBNxiii \undefined \def \showISBNxiii  #1{\unskip}     \fi
\ifx \showISSN     \undefined \def \showISSN      #1{\unskip}     \fi
\ifx \showLCCN     \undefined \def \showLCCN      #1{\unskip}     \fi
\ifx \shownote     \undefined \def \shownote      #1{#1}          \fi
\ifx \showarticletitle \undefined \def \showarticletitle #1{#1}   \fi
\ifx \showURL      \undefined \def \showURL       {\relax}        \fi
\providecommand\bibfield[2]{#2}
\providecommand\bibinfo[2]{#2}
\providecommand\natexlab[1]{#1}
\providecommand\showeprint[2][]{arXiv:#2}

\bibitem[\protect\citeauthoryear{Baziotis, Kang, and Mendis}{Baziotis et~al\mbox{.}}{2024}]%
        {DIAS}
\bibfield{author}{\bibinfo{person}{Stefanos Baziotis}, \bibinfo{person}{Daniel Kang}, {and} \bibinfo{person}{Charith Mendis}.} \bibinfo{year}{2024}\natexlab{}.
\newblock \showarticletitle{Dias: Dynamic Rewriting of {Pandas} Code}.
\newblock \bibinfo{journal}{\emph{Procs. of ACM on Management of Data (PACMMOD)}} \bibinfo{volume}{2}, \bibinfo{number}{1} (\bibinfo{date}{Feb.} \bibinfo{year}{2024}), 27.
\newblock
\urldef\tempurl%
\url{https://doi.org/10.1145/3639313}
\showDOI{\tempurl}


\bibitem[\protect\citeauthoryear{Contributors}{Contributors}{2012}]%
        {Soot}
\bibfield{author}{\bibinfo{person}{Soot Contributors}.} \bibinfo{year}{2012}\natexlab{}.
\newblock \bibinfo{title}{Soot: a Java Optimization Framework}.
\newblock \bibinfo{howpublished}{\url{https://www.sable.mcgill.ca/soot/}}.
\newblock


\bibitem[\protect\citeauthoryear{Emani, Ramachandra, Bhattacharya, and Sudarshan}{Emani et~al\mbox{.}}{2016}]%
        {equivsql2016}
\bibfield{author}{\bibinfo{person}{K.~Venkatesh Emani}, \bibinfo{person}{Karthik Ramachandra}, \bibinfo{person}{Subhro Bhattacharya}, {and} \bibinfo{person}{S. Sudarshan}.} \bibinfo{year}{2016}\natexlab{}.
\newblock \showarticletitle{Extracting Equivalent SQL from Imperative Code in Database Applications}. In \bibinfo{booktitle}{\emph{Procs. of SIGMOD}}. \bibinfo{pages}{1781–1796}.
\newblock


\bibitem[\protect\citeauthoryear{et~al.}{et~al.}{2017}]%
        {weld}
\bibfield{author}{\bibinfo{person}{Shoumik~Palkar et al.}} \bibinfo{year}{2017}\natexlab{}.
\newblock \showarticletitle{Weld: A Common Runtime for High Performance Data Analytics}. In \bibinfo{booktitle}{\emph{CIDR ’17}}.
\newblock


\bibitem[\protect\citeauthoryear{Hecht and Ullman}{Hecht and Ullman}{1972}]%
        {hechtullman1972}
\bibfield{author}{\bibinfo{person}{M.~S. Hecht} {and} \bibinfo{person}{J.~D. Ullman}.} \bibinfo{year}{1972}\natexlab{}.
\newblock \showarticletitle{Flow graph reducibility}. In \bibinfo{booktitle}{\emph{Procs. ACM Symp. Theory of Computation {(STOC)}}}. \bibinfo{pages}{238–250}.
\newblock


\bibitem[\protect\citeauthoryear{Jindal et~al\mbox{.}}{Jindal et~al\mbox{.}}{2021}]%
        {Magpie}
\bibfield{author}{\bibinfo{person}{Alekh Jindal} {et~al\mbox{.}}} \bibinfo{year}{2021}\natexlab{}.
\newblock \showarticletitle{Magpie: Python at Speed and Scale using Cloud Backends}. In \bibinfo{booktitle}{\emph{Conf. on Innovative Data Systems Research (CIDR)}}.
\newblock


\bibitem[\protect\citeauthoryear{Khedker, Sanyal, and Karkare}{Khedker et~al\mbox{.}}{2009}]%
        {dfabook}
\bibfield{author}{\bibinfo{person}{Uday Khedker}, \bibinfo{person}{Amitabha Sanyal}, {and} \bibinfo{person}{Bageshri Karkare}.} \bibinfo{year}{2009}\natexlab{}.
\newblock \bibinfo{booktitle}{\emph{Data Flow Analysis: Theory and Practice}}.
\newblock \bibinfo{publisher}{CRC Press, Inc.}
\newblock


\bibitem[\protect\citeauthoryear{Modin}{Modin}{2024}]%
        {modin}
Modin \bibinfo{year}{2024}\natexlab{}.
\newblock \bibinfo{howpublished}{\url{https://github.com/modin-project/modin}}.
\newblock


\bibitem[\protect\citeauthoryear{Morton et~al\mbox{.}}{Morton et~al\mbox{.}}{2020}]%
        {delayrelay}
\bibfield{author}{\bibinfo{person}{John~Magnus Morton} {et~al\mbox{.}}} \bibinfo{year}{2020}\natexlab{}.
\newblock \showarticletitle{DelayRepay: Delayed Execution for Kernel Fusion in Python}. In \bibinfo{booktitle}{\emph{ACM SIGPLAN Intl Symp.\ on Dynamic Languages (DLS)}}. \bibinfo{publisher}{ACM}.
\newblock


\bibitem[\protect\citeauthoryear{Muchnick}{Muchnick}{1997}]%
        {muchnik1997}
\bibfield{author}{\bibinfo{person}{S Muchnick}.} \bibinfo{year}{1997}\natexlab{}.
\newblock \bibinfo{booktitle}{\emph{Advanced Compiler Design Implementation}}.
\newblock \bibinfo{publisher}{Morgan Kaufmann}.
\newblock


\bibitem[\protect\citeauthoryear{Petersohn, Macke, Xin, Ma, Lee, Mo, Gonzalez, Hellerstein, Joseph, and Parameswaran}{Petersohn et~al\mbox{.}}{2020}]%
        {tsds}
\bibfield{author}{\bibinfo{person}{Devin Petersohn}, \bibinfo{person}{Stephen Macke}, \bibinfo{person}{Doris Xin}, \bibinfo{person}{William Ma}, \bibinfo{person}{Doris Lee}, \bibinfo{person}{Xiangxi Mo}, \bibinfo{person}{Joseph~E. Gonzalez}, \bibinfo{person}{Joseph~M. Hellerstein}, \bibinfo{person}{Anthony~D. Joseph}, {and} \bibinfo{person}{Aditya Parameswaran}.} \bibinfo{year}{2020}\natexlab{}.
\newblock \showarticletitle{Towards Scalable Dataframe Systems}.
\newblock \bibinfo{journal}{\emph{Proc. VLDB Endow.}} \bibinfo{volume}{13}, \bibinfo{number}{12} (\bibinfo{date}{July} \bibinfo{year}{2020}), \bibinfo{pages}{2033–2046}.
\newblock


\bibitem[\protect\citeauthoryear{PySpark}{PySpark}{2024}]%
        {pysparkpandas}
PySpark \bibinfo{year}{2024}\natexlab{}.
\newblock \bibinfo{title}{Pandas {API} on {Spark}}.
\newblock \bibinfo{howpublished}{\url{https://spark.apache.org/docs/latest/api/python/user_guide/pandas_on_spark}}.
\newblock
\newblock
\shownote{Accessed Sep 2024.}


\bibitem[\protect\citeauthoryear{Ray}{Ray}{2024}]%
        {ray}
Ray \bibinfo{year}{2024}\natexlab{}.
\newblock \bibinfo{title}{Ray: a unified framework for scaling {AI} and {Python} applications}.
\newblock \bibinfo{howpublished}{\url{https://github.com/ray-project/ray}}.
\newblock
\newblock
\shownote{Accessed Sep 2024.}


\bibitem[\protect\citeauthoryear{Rocklin}{Rocklin}{2015}]%
        {rocklin2015dask}
\bibfield{author}{\bibinfo{person}{Matthew Rocklin}.} \bibinfo{year}{2015}\natexlab{}.
\newblock \showarticletitle{Dask: Parallel computation with blocked algorithms and task scheduling}. In \bibinfo{booktitle}{\emph{Proceedings of the 14th {Python} in science conference}}. Citeseer.
\newblock


\bibitem[\protect\citeauthoryear{Singh, Sahu, and Sudarshan}{Singh et~al\mbox{.}}{2021}]%
        {self_paper}
\bibfield{author}{\bibinfo{person}{Bhushan~Pal Singh}, \bibinfo{person}{Mudra Sahu}, {and} \bibinfo{person}{S. Sudarshan}.} \bibinfo{year}{2021}\natexlab{}.
\newblock \showarticletitle{Optimizing Data Science Applications using Static Analysis}. In \bibinfo{booktitle}{\emph{Int'l Symp. on Database Prog. Languages (DBPL)}}. \bibinfo{pages}{23–27}.
\newblock


\bibitem[\protect\citeauthoryear{Yan, Lin, and He}{Yan et~al\mbox{.}}{2023}]%
        {MagicPush}
\bibfield{author}{\bibinfo{person}{Cong Yan}, \bibinfo{person}{Yin Lin}, {and} \bibinfo{person}{Yeye He}.} \bibinfo{year}{2023}\natexlab{}.
\newblock \showarticletitle{Predicate Pushdown for Data Science Pipelines}.
\newblock \bibinfo{journal}{\emph{Proc. ACM Manag. Data}} \bibinfo{volume}{1}, \bibinfo{number}{2} (\bibinfo{date}{jun} \bibinfo{year}{2023}), 28.
\newblock


\bibitem[\protect\citeauthoryear{Zhang and Shen}{Zhang and Shen}{2021}]%
        {cun}
\bibfield{author}{\bibinfo{person}{Guoqiang Zhang} {and} \bibinfo{person}{Xipeng Shen}.} \bibinfo{year}{2021}\natexlab{}.
\newblock \showarticletitle{{Best-Effort Lazy Evaluation for Python Software Built on APIs}}. In \bibinfo{booktitle}{\emph{ECOOP 2021}}.
\newblock


\end{thebibliography}

\fullversion{
\begin{appendices}
\newpage

\section{Remaining Algorithms }
In this appendix, we have added detailed algorithms. 

\begin{algorithm}[H]
\caption{Check for Row Selection statement}
\label{algo:is_row_selection}
\begin{algorithmic}
\Procedure{\textbf{is\_row\_selection}}{task\_graph \textbf{u}}

\If {(operation(u) is 'get-item') \\
    \quad\quad\quad \textbf{and} (u has 1 child) \\ 
    \quad\quad\quad \textbf{and} (operator\_of(child\_of(u))) in $[>, >=, <= <, ==, !=, \&, |]$}

    \State \Return True
    \EndIf
    
    \State \Return False
\EndProcedure
\end{algorithmic}
\end{algorithm}

\begin{algorithm}[H]
\caption{Check for any merge conflict during row\_selection}
\label{algo:check_merge_conflict}
\begin{algorithmic}
\Procedure{\textbf{has\_merge\_conflict}}{task\_graph \textbf{u}, task\_graph \textbf{v}}

    \LineComment {Both u and v are row\_selection statements}
        \LineComment {u preceeds v in task graph (program start...u-v...end)}
  \If {(any operators(v) is in [Agg, complete dataset op])}
    \State \Return True

    \LineComment {No conflicts. Can be merged together}
    \EndIf
  
  \State \Return False

\EndProcedure
\end{algorithmic}
\end{algorithm}

\begin{algorithm}[H]
\caption{Extended row\_selection}
\label{algo:extended_row_selection}
\begin{algorithmic}
\Procedure{\textbf{Row\_selection\_extended}}{task\_graph \textbf{G}}

  \ForAll {(node \textit{v} in post-dfs order of task graph G)}

    \While {(True)}

    \State $u \gets $ Immediate preceding operation of v in task graph
    
    \LineComment {u preceeds v in task graph (program start...u-v...end)}

     \If { (v is a row\_selection statement) \\
        \quad\quad\quad\quad\quad\quad\textbf{and} (u is not a row\_selection statement) \\
        \quad\quad\quad\quad\quad\quad\textbf{and} (has\_swap\_conflict(u,v) is \textbf{False}) \\
        \quad\quad\quad\quad\quad\quad\textbf{and} (v is live at all program paths from u to  end)}
        
       \State Remove v from all other paths from u to  end
       \State Swap(u,v) and \textbf{continue}
    \EndIf
    \\
     \If { (v is a row\_selection statement) \\
        \quad\quad\quad\quad\quad\quad\textbf{and} (u is not a row\_selection statement) \\
        \quad\quad\quad\quad\quad\quad\textbf{and} (has\_swap\_conflict(u,v) is \textbf{True}) \\
        \quad\quad\quad\quad\quad\quad\textbf{and} (v is live at all program paths from u to  end)}
        
       \State Consider (u,v) as \textbf{compound\_row\_selection}  and \textbf{continue}

    \EndIf
    \\

     \If { (v is a compound\_row\_selection statement) \\
        \quad\quad\quad\quad\quad\quad\textbf{and} (has\_swap\_conflict(u,v) is \textbf{False}) \\
        \quad\quad\quad\quad\quad\quad\textbf{and} (v is live at all program paths from u to  end) \\
        \quad\quad\quad\quad\quad\quad\textbf{and} ( (u is not a row\_selection statement) \\ \quad\quad\quad\quad\quad\quad\textbf{OR} (has\_merge\_conflicts(u, get\_row\_selection\_of(v)) is \textbf{False}))}
        
        \State Swap(u,v) and \textbf{continue}
    \Else \If {(v is a compound\_row\_selection statement) \\
        \quad\quad\quad\quad\quad\quad\quad\textbf{and} (v is live at all program paths from u to  end)}
           \State Consider (u,v) as \textbf{compound\_row\_selection} and \textbf{continue}
    
    \EndIf
    \EndIf

    \EndWhile
    \EndFor

\EndProcedure
\end{algorithmic}
\end{algorithm}

\end{appendices}
}



\end{document}